\documentclass[aps,prb,twocolumn,superscriptaddress,
showpacs,floatfix]{revtex4-1}
\usepackage[caption=false]{subfig}
\usepackage{graphicx}
\usepackage{dcolumn}
\usepackage{bm}
\usepackage{calc} 
\usepackage{amsfonts} 
\usepackage{amssymb} 
\usepackage{amsmath}
\usepackage{subfig} 
\usepackage{natbib}
\usepackage{array,color}
\usepackage{amsbsy}

\begin{document}

\title{Electrical transport near   quantum criticality in 
low dimensional organic superconductors}

\author{M. Shahbazi} 
\email{maryam.shahbazi@usherbrooke.ca}
\author{C. Bourbonnais}
\email{claude.bourbonnais@USherbrooke.ca }
\affiliation{%
Regroupement Qu\'ebecois sur les Mat\'eriaux de Pointe, D\'epartement de physique, Universit\'e de Sherbrooke, Sherbrooke, Qu\'ebec, Canada, J1K-2R1}%

\date{\today}

\begin{abstract}

We propose a theory of longitudinal resistivity in the normal  phase of quasi-one-dimensional organic superconductors  near the quantum critical point where antiferromagnetism borders with superconductivity under pressure.  The linearized semi-classical Boltzmann equation is solved numerically, fed in by the half-filling electronic  umklapp scattering vertex as derived from one-loop  renormalization group calculations for the quasi-one-dimensional electron gas model. The momentum and temperature dependence of umklapp scattering has an important impact on the behaviour of longitudinal resistivity in the the normal phase. Resistivity is found to be linear  in temperature around the quantum critical point at which spin-density-wave order joins superconductivity along the antinesting axis, to gradually evolve  towards the Fermi liquid behaviour in the limit of weak superconductivity. A comparison is made between theory and   experiments performed on the (TMTSF)$_2$PF$_6$ member of the Bechgaard salt series  under pressure.
\end{abstract}

\maketitle
\section{Introduction}

The quasi-one-dimensional (quasi-1D) organic conductors (TMTSF)$_2$X, 
   called  the  Bechgaard salts,  
  are  known to be among the  first  examples of correlated electron systems in which superconductivity (SC) borders with antiferromagnetism or spin-density-wave order (SDW) in their phase diagram \cite{Jerome80,Jerome82,Bourbon08,Brown15}. Both  orderings are brought in close proximity by tuning pressure which can be achieved either hydrostatically or by chemical means. This  results in  a characteristic SDW-SC sequence of instabilities that  shares  many common traits with other  unconventional superconductors including members of  cuprate \cite{Taillefer10,Armitage10}, heavy-fermion \cite{Tomita15,Scalapino12}, and  pnictide \cite{Stewart11,Scalapino12} series of compounds. This sequence  is typically preceded by  a metallic phase  with unusual properties.   For the Bechgaard salts, an   anomalous  metallic phase occurs nearby the critical pressure $P_c$, as a quantum critical point (QCP)  connecting the ordering temperature $T_c$ for SC at its optimal value with the vanishing $T_{\rm SDW}$ for SDW\cite{Brown15,Bourbon08}. The temperature dependence of longitudinal resistivity near this point is found to depart  from the  $T^2$ behaviour for a    Fermi liquid. A detailed analysis has revealed that  resistivity is instead  linear in temperature close to $P_c$, as commonly found near a QCP\cite{Taillefer10}. It smoothly evolves toward the Fermi liquid prediction as pressure is tuned away from $P_c$, in apparent  correspondence  with the gradual suppression of $T_c$ under pressure \cite{DoironLeyraud09,DoironLeyraud10}. 

  Generally linked to a  QCP  with  SDW is the presence of spin fluctuations in the metallic state. These fluctuations may give rise to an important source of electron-electron umklapp scattering. As a mechanism of momentum dissipation, umklapp enters as a key determinant in the temperature dependence of  resistivity\cite{Gorkov98,Zheleznyak99}. In systems like the Bechgaard salts, this scattering process is primarily the consequence of a weak dimerization of the organic stacks which imparts a half-filled character to the electron  band and allows the longitudinal transfer of charge carriers across the Fermi surface \cite{Emery82}. The  existence of SDW fluctuations in the metallic phase of these materials   has been amply borne out by NMR measurements\cite{Bourbon84,Creuzet87b,Wu05,Brown08,Kimura11}. The temperature dependence  of the   nuclear spin relaxation rate $T_1^{-1}$  shows a pronounced    enhancement of the Curie-Weiss form,  at variance with  the  linear$-T$ Korringa law expected  for a Fermi liquid. The amplitude of enhancement  is in precise correspondence with the anomalous  resistivity  and the size of  $T_c$  over the broad range of pressure where superconductivity is present\cite{Bourbon09,Kimura11}, showing an intimate connection between SDW and Cooper pairing.  
  
  From a theoretical point of view,  one-loop renormalization group (RG) studies on  the quasi-1D electron gas model with umklapp scattering have previously shown how the  SDW-SC sequence of instabilities  can be reproduced by varying the transverse next-to-nearest-neighbour hopping term $t'_\perp$, which acts as  an antinesting term that simulates the role of pressure in the model\cite{Nickel06,Bourbon09}. Along the antinesting axis, a critical value $t_\perp'^*$ can be defined   as the analogue of the actual QCP at $P_c$. This is  where  $T_{\rm SDW}$ is suppressed and a d-wave  SC  (SCd) instability appears at a maximum $T_c$. The SDW-SCd pattern is also found to be accompanied by  SDW fluctuations over a wide  temperature domain  of the non ordered  state. Above $t_\perp'^*$ for instance,  the calculated SDW susceptibility in the normal phase fits a Curie-Weiss form in agreement with the one extracted from NMR experiments. This enhancement of SDW correlations is mainly ascribed  to the positive response of  umklapp scattering to the growth of  d-wave pairing\cite{Bourbon09,Sedeki12}.

From the one-loop electron-electron vertex functions obtained  for the electron gas, the quasi-particle scattering rate has been computed from the self-energy\cite{Sedeki12,DoironLeyraud10}. Near $t_\perp'^*$, the rate  develops a linear $T$ dependence above $T_c$, to which a Fermi liquid $T^2$ component subsequently  adds and grows as $t_\perp'$ distances from the QCP and $T_c$ decreases. This   qualitatively agrees with the polynomial $T$-analysis of resistivity data in systems like (TMTSF)$_2$PF$_6$ and (TMTSF)$_2$ClO$_4$ over the whole pressure domain where superconductivity is found \cite{DoironLeyraud09,DoironLeyraud10}. An alternative  self-energy approach to  resistivity data  has been proposed  from a hot-cold spots picture of the SDW nesting along  quasi-1D  Fermi surface in the conventional framework  of the QCP where  superconductivity is neglected\cite{Meier13}.

In this work we intend to push the calculation of longitudinal resistivity a step beyond   the  self-energy  approach by solving the semi-classical Boltzmann equation. This is achieved numerically with the aid of the RG method which is employed  to compute the momentum and temperature dependent umklapp scattering vertex function   entering  the collision term of the Boltzmann equation.  Our  procedure is similar in the outline  with the  one recently proposed  by Buhmann {\it et al.,}\cite{Buhmann13} in the study of  transport properties of the hole-doped cuprate superconductors.

The results put forward below show that resistivity develops a metallic linear-$T$ dependence
 as antinesting approaches its critical value $t_\perp'^*$  where SDW and SCd orders meet. This quantum critical behaviour in resistivity ensues from the anisotropic growth of half-filling umklapp scattering with lowering temperature,  a consequence of  reinforcement of commensurate SDW correlations by  SCd pairing in the metallic phase, which persists down to $T_c$. Above $t_\perp'^*$, the  reinforcement undergoes a gradual decline that defines   an extended region of quantum criticality where the temperature dependence of resistivity can  fit  a power law dependence $\rho(T) \sim T^\alpha$  with an  exponent $\alpha < 2$. The exponent approaches the Fermi liquid limit  $\alpha\simeq2$ at  low enough temperature and for  sufficiently large $t_\perp'$, namely where $T_c$ becomes small.  Anisotropy developed by umklapp scattering  is found to have   a sizeable  impact on the  momentum   dependence of the scattering rate extracted from  resistivity.  

The linearized Boltzmann transport theory is introduced in  Sec.~II and  Appendix A. In  Sec.~III and Appendix B, we present the  RG approach of the quasi-1D electron gas,  from which the results for the phase diagram along with the  temperature and momentum dependence of umklapp scattering vertex entering  the Boltzmann equation are given. In Sec.~IV, numerical results for resistivity and their analysis in terms of renormalized umklapp vertex function are detailed.  A comparison with existing experiments performed on the (TMTSF)$_2$PF$_6$ member of the Bechgaard salts is presented. We summarize and conclude this work in Sec.~V.   
 
\section{Boltzmann equation}
The semicalssical Boltzmann equation describes the variation of the  quasi-particle Fermi  distribution function $f$ due to collisions and external forces. In the presence of a spatially uniform and static electric field $\boldsymbol{\cal E}$  coming from a  one-body electrostatic potential, the Boltzmann equation  reduces to the expression 
\begin{align}\label{eq1}
\dfrac{df(\bm{k})}{dt} &= e\,\mathbf{\cal E}\cdot {\nabla}_{\hbar \bm{k}}f 
=  \left[\dfrac{\partial f(\bm{k})}{\partial t}\right]_{\text{coll}},
\end{align} 
where $e$ is the electric charge. The collision integral for an array of $N_P$ chains of length $L$  takes the form
\begin{align}\label{eq2}
\left[\dfrac{\partial f(\bm{k})}{\partial t}\right]_{\text{coll}}& = \   (LN_P)^{-2} \sum \limits_{\bm{k}_2,\bm{k}_3,\bm{k}_4}w(\bm{k},\bm{k}_2;\bm{k}_3,\bm{k}_4)\cr
 & \times  \{{ f(\bm{k})f(\bm{k}_2)[1-f(\bm{k}_3)][1-f(\bm{k}_4)]}\cr
&- [1-f(\bm{k})][1-f(\bm{k}_2)]f(\bm{k}_3)f(\bm{k}_4)\},
\end{align}
for scattering in and out processes. The electron-electron (longitudinal umklapp) contribution to scattering rate $w$ is obtained from the  Fermi golden rule,
\begin{align}\label{eq3}
w(\bm{k},\bm{k}_2;& \bm{k}_3,\bm{k}_4)= {1\over 2}{\vert\langle\bm{k},\bm{k}_2\vert g_3 \vert\bm{k}_3,\bm{k}_4\rangle -\langle\bm{k},\bm{k}_2\vert g_3 \vert\bm{k}_4,\bm{k}_3\rangle\vert}^2 \cr
&\times  \delta_{\bm{k}+\bm{k}_2,\bm{k}_3+\bm{k}_4+ \bm{G}}
{2\pi \over \hbar}\delta(\varepsilon_{\bm{k}}+\varepsilon_{\bm{k}_2}-\varepsilon_{\bm{k}_3}-\varepsilon_{\bm{k}_4}) \cr
\end{align}
where $\bm{G}=(4k_F,0) $ is the longitudinal reciprocal lattice vector for half-filling umklapp scattering, and $k_F=\pi/(2a)$ is the longitudinal (1D) Fermi wave vector for   dimerized chains.

To linearize the Boltzmann equation, we introduce the normalized deviation function $\phi_{\bm{k}}$ in the Fermi distribution function\cite{Haug08},
\begin{eqnarray}
 \label{eq5}
f(\bm{k})=\dfrac{1}{e^{\beta \epsilon^{p}_{\bm{k}}-\phi_{\bm{k}}+1}},
\end{eqnarray}
where $\beta= 1/T$ $(k_B=1)$. The electron energy spectrum is the one of  the quasi-1D electron gas model
\begin{align}
\epsilon^{p}_{\bm{k}} =& \hbar v_F(pk-k_F)   + \epsilon_\perp(k_\perp).
\label{Ek}
\end{align}
The spectrum  comprises a  longitudinal part   linearized around the right/left 1D Fermi points $pk_F=\pm k_F$  with $v_F$ as the longitudinal Fermi velocity. The transverse part is given by
\begin{equation}
\label{ }
\epsilon_\perp(k_\perp)= -2t_\perp\cos k_\perp d_\perp -2t_{\perp}^\prime \cos 2k_\perp d_\perp, 
\end{equation}
 where $t_\perp$ and $t_\perp'$ are the first and second nearest-neighbour   interchain hopping terms and $d_\perp$ is the interchain distance.

The first order expansion $f(\bm{k})\simeq f^0(\bm{k})+f^0(\bm{k})[1-f^0(\bm{k})]\phi_{\bm{k}}$ of the distribution function  yields after substitution in   (\ref{eq1}),
\begin{eqnarray}
\label{BE}
\mathcal{L}\phi_{\bm{k}} = e\beta\mathbf{\cal E}\cdot v_{\boldsymbol{k}},
\end{eqnarray}
 where $\mathbf{\cal E}\sim {\cal O}(\phi)$. This is the linearized Boltzmann equation in which  the collision operator $\mathcal{L}$  obeys the relation
\begin{eqnarray}\label{eq8}
\mathcal{L} \phi_{\bm{k}} =  \sum\limits_{\bm{k^\prime}} \mathcal{L}_{\bm{k},\bm{k^\prime}}\phi_{\bm{k^\prime}}.
\end{eqnarray}
The matrix elements are of the form
\begin{widetext}
\begin{eqnarray}
\label{L1}
\mathcal{L}_{\bm{k},\bm{k^\prime}} = \dfrac{1}{{(L N_P)}^2} &\sum\limits_{\bm{k}_2,\bm{k}_3,\bm{k}_4}& {1\over 2}{\vert\langle\bm{k},\bm{k}_2\vert g_3 \vert\bm{k}_3,\bm{k}_4\rangle -\langle\bm{k},\bm{k}_2\vert g_3 \vert\bm{k}_4,\bm{k}_3\rangle\vert}^2
 \frac{2\pi}{\hbar}  \delta_{\bm{k}+\bm{k}_2,\bm{k}_3+\bm{k}_4+ p \bm{G}} \delta(\varepsilon^p_{\bm{k}}+\varepsilon^{p_2}_{\bm{k}_2}-\varepsilon^{p_3}_{\bm{k}_3}-\varepsilon^{p_4}_{\bm{k}_4}) \nonumber \\
&\times & \dfrac{f^0(\bm{k}_2)[1-f^0(\bm{k}_3)][1-f^0(\bm{k}_4)]}{[1-f^0(\bm{k})]} (\delta_{\bm{k},\bm{k^\prime}} + \delta_{\bm{k}_2,\bm{k^\prime}} - \delta_{\bm{k}_3,\bm{k^\prime}} - \delta_{\bm{k}_4,\bm{k^\prime}}) = \sum_{i=1}^4 \mathcal{L}^{[i]}_{\bm{k},\bm{k^\prime}},
\end{eqnarray}
\end{widetext}
which separates  into a diagonal ($\mathcal{L}^{[1]}$) and three off-diagonal terms $(\mathcal{L}^{[2-4]}$).  In the Appendix A, each term is evaluated explicitly in the framework of the quasi-1D electron gas model introduced in Sec.~\ref{RG}.

For an electric field $\boldsymbol{\cal E}= {\cal E}_a\hat{a}$ oriented along the chain   direction, the  corresponding electrical current density $j_a=\sigma_{a}{\cal E}_a$ allows us to extract the  conductivity $\sigma_{a}$, as the inverse of longitudinal resistivity $\rho_{a}$. To first order in $\phi$, the current density is given by  
\begin{align}\label{eq11}
{j}_a &= \frac{2e}{L N_P d_\perp}\sum\limits_{\bm{k}}v_{F}f(\bm{k}) \nonumber \\
&\simeq \frac{2e}{L N_P d_\perp}\sum\limits_{\bm{k}}v_{F}f^0(\mathbf{k})[1-f^0(\bm{k})]\phi_{k_\perp}.
\end{align}
Since for all temperatures of interest, the product $f^0[1-f^0]$ is strongly peaked at the Fermi level,   the deviation $\phi_{\boldsymbol{k}}\to \phi_{\boldsymbol{k}^p_F} \equiv \phi_{k_\perp} $ can be evaluated  at the Fermi  wave vector  $\boldsymbol{k}^p_F=(k^p_F(k_\perp),k_\perp)$, whose location on the $p$ Fermi surface sheet   is entirely parametrized by  $k_\perp$ from the equation $ \epsilon^{p}_{ \bm{k}_F^p}=0$. By introducing the normalized deviation $\bar{\phi}_{k_\perp} = {\phi}_{k_\perp}/{(  \beta e {\cal E}_ad_\perp)}$ and after   an energy integration following (\ref{sum}),
the  2D conductivity or the inverse   resistivity in the longitudinal direction becomes  
\begin{eqnarray}
\label{sigma}
\sigma_{a}= \rho_{a}^{-1}=\frac{e^2}{\hbar} \langle \bar{\phi}_{k_\perp}\rangle_{\rm FS},
\end{eqnarray}
where $\langle \bar{\phi}_{k_\perp}\rangle_{\rm FS} =  {1}/{N_P}\sum_{k_{\perp}}\bar{\phi}_{k_{\perp}}$
is the  deviation function averaged over the Fermi surface. The function $\bar{\phi}_{k_{\perp}}$ satisfies the equation  
\begin{equation}
\label{LinBE}
\sum_{i,k_\perp'}\bar{\mathcal{L}}^{[i]}_{k_\perp,k_\perp'}\bar{\phi}_{k_\perp'}=1,
\end{equation}
whose explicit  expression is given in (\ref{LBEb}) and for which we have introduced the dimensionless operator ${\bar{\mathcal{L}}^{[i]} =  \frac{\pi d_\perp}{v_F}\mathcal{L}^{[i]}}$. The above expression is fed by the momentum and temperature  dependence of the umklapp vertex function. This function is  provided by the RG approach to the quasi-1D electron gas model that is  introduced next.

\section{ Renormalized umklapp scattering for the electron gas model}
\label{RG}
\subsection{The model}
The quasi-1D electron gas model in its  standard form comprises, besides its one-electron energy spectrum (\ref{Ek}), three electron-electron coupling constants that will be defined here on the 
Fermi surface sheets $\boldsymbol{k}_F^p$. These are the backward and the forward scattering amplitudes $g_1(\boldsymbol{k}_{F,1}^-,\boldsymbol{k}_{F,2}^+;\boldsymbol{k}_{F,3}^-,\boldsymbol{k}_{F,4}^+)$ and $g_2(\boldsymbol{k}_{F,1}^+,\boldsymbol{k}_{F,2}^-;\boldsymbol{k}_{F,3}^-,\boldsymbol{k}_{F,4}^+)$ for normal scattering processes between right and left moving carriers, and $g_3(\boldsymbol{k}_{F,1}^p,\boldsymbol{k}_{F,2}^p;\boldsymbol{k}_{F,3}^{-p},\boldsymbol{k}_{F,4}^{-p})$ for umklapp scattering along the chains,   whose momentum conservation involves the reciprocal lattice vector $\boldsymbol{G}=(4k_F,0)$ at half-filling. These coupling constants can be seen as phenomenological parameters of the model whose range at the bare level    can be  fixed from experimental data. We will follow \cite{Bourbon09,Sedeki12,Nickel06}, and use the  typical   values (normalized by  $\hbar\pi v_F$), $g_1\simeq 0.32$ and $g_2 \simeq 0.64$ for the two normal processes, consistent with the observed enhancement of uniform magnetic susceptibility\cite{Wzietek93} and the scale of $T_{\rm SDW}$ at low pressure\cite{Klemme95,Moser98}. As for the  umklapp term, its bare value is primarily proportional to the     dimerization gap $\Delta_D$ \cite{Grant83} of the organic stacks ($g_3\approx g_1\Delta_D/E_F$)\cite{Barisic81,Mila94}, which is small in the Bechgaard salts. Although rather weak at the bare level, umklapp processes have a strong influence on spin fluctuations at low energy and enter as  a key determinant in the temperature dependence of longitudinal resistivity as we will see. The bare umklapp amplitude will thus be varied in the interval  $g_3\simeq 0.02\ldots 0.035$, in accordance with the range of values expected for $\Delta_D$\cite{Grant83}. For the model spectrum (\ref{Ek}), we  shall use  $E_F= 3000$K and $t_\perp=200$K for the Fermi energy and transverse hopping, as representative figures of the band structure for the Bechgaard salts\cite{Grant82,Ducasse85}. The transverse second nearest-neighbour hopping $t_\perp'$, which  introduces nesting alterations, will be varied continuously to simulate pressure effects.   

We follow the lines of previous works and apply the Kadanoff-Wilson RG approach to the quasi-1D electron gas model which is outlined in Appendix~B\cite{Bourbon91,Bourbon03,Nickel06,Sedeki12}. Each constant energy surface is first divided  into  ${N_P(=60)}$ patches, each centered  on a discrete  value  of $k_\perp$ parametrizing each Fermi sheet $\boldsymbol{k}_F^p=(k^p_F(k_\perp),k_\perp)$.  The successive integration of electronic degrees of freedom as a function of the energy distance  from the Fermi surface  comprises perturbative contributions to the scattering amplitudes  coming from closed loops, vertex corrections and ladder diagrams of the density-wave (Peierls) and electron-electron (Cooper) scattering channels. The resulting RG flow equations for the $k_\perp$-dependent scattering  amplitudes $g_{1,2,3}(k_{\perp1},k_{\perp2};k_{\perp3},k_{\perp4})$ on the Fermi surface are  reproduced in (\ref{Flowg}) of  Appendix~B. Their $k_\perp$ dependence takes into account the transverse momentum variables in both  Peierls and Cooper channels, along with  the longitudinal momentum transfer in each channel  which is adjusted so that the momentum of ingoing and outgoing particles fall on the warped Fermi surface. 

These flow equations are completed by those of  the susceptibilities $\chi_\mu(\boldsymbol{q}_0)$ given in (\ref{FlowKi}). These serve to signal an instability against either $\mu=$SDW or $\mu=$SCd types of ordering at the corresponding wave vectors $\boldsymbol{q}_0=(2k_F,\pi)$ and $\boldsymbol{q}_0=0$.
\subsection{The phase diagram} 
\begin{figure}
\includegraphics[width=8cm]{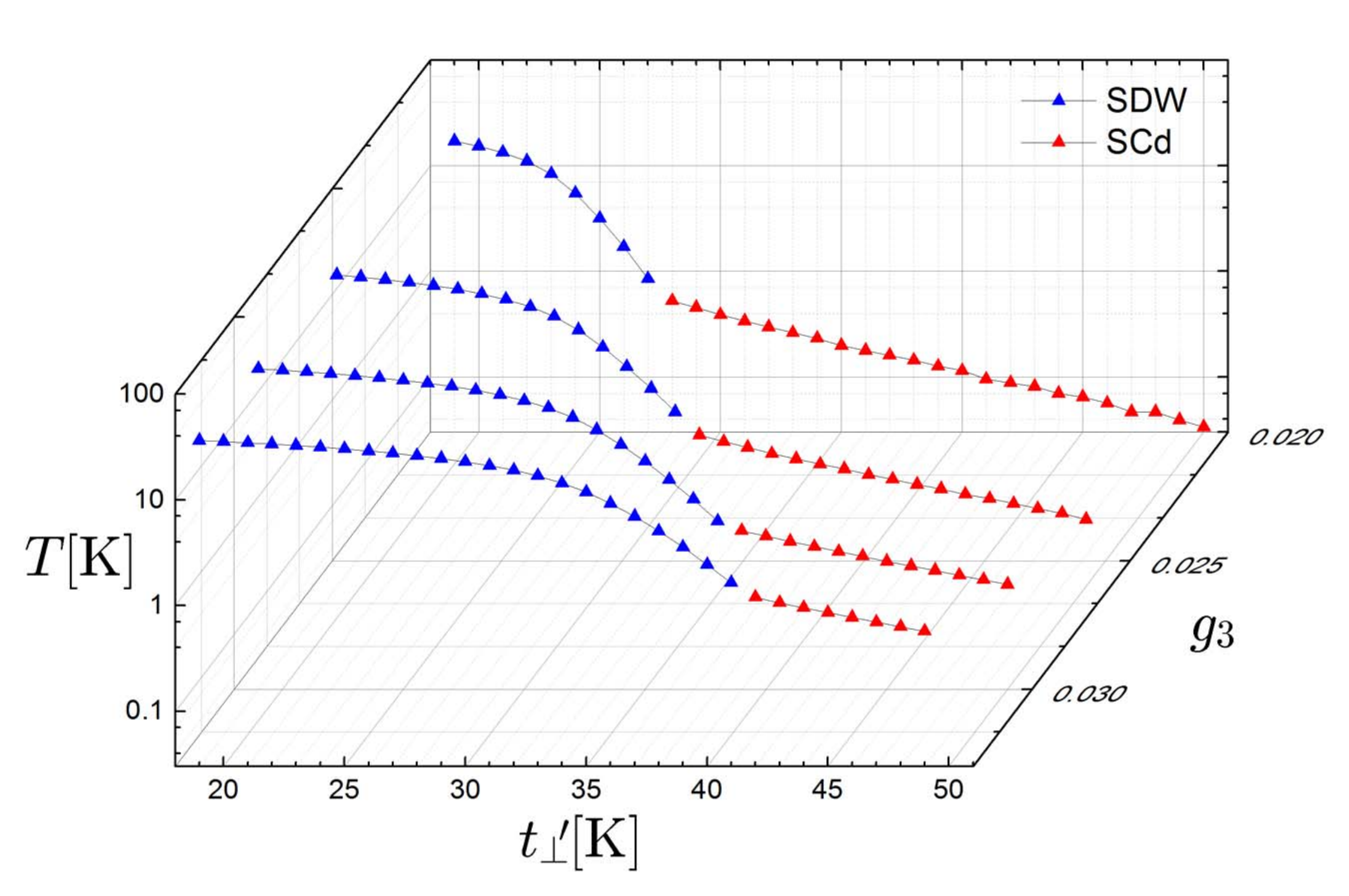}
\caption{ (Color online) Calculated phase diagram of the quasi-one-dimensional electron gas model  showing the SDW-SCd sequence of instabilities as a function of antinesting  $t_\perp'$ and  initial $g_3$.}
\label{Phases}
\end{figure}

 The integration of the flow equations for the scattering amplitudes (\ref{Flowg}) and susceptibilities (\ref{FlowKi}-\ref{FlowKiz})  leads to their corresponding renormalized values at   temperature $T$. For the above set of model parameters, the singularities in $\chi_{\rm SDW}(2k_F,\pi)$ and   $\chi_{\rm SCd}(0)$ at temperatures $T_{\rm SDW}$ and 
 $T_{\rm SCd}$ enable to follow the SDW-SCd sequence  of instabilities  as a function of the antinesting parameter $t_\perp'$ and    initial   $g_3$. The resulting phase diagrams  are shown in Fig.~\ref{Phases}. The variation of $T_{\rm SDW}$ shows a characteristic monotonic decrease with $t_\perp'$ up to the approach of the critical value  $t_\perp'^*$ where  $T_{\rm SDW}$ undergoes a  rapid  decrease, followed  by the emergence of a   SCd instability. The latter  reaches its maximum $T_c$ at   $t_\perp'^*$, followed  by its steady decrease for larger $t_\perp'$. The impact of increasing the initial $g_3$ on  this sequence is to magnify antiferromagnetism and $T_{\rm  SDW}$, which ultimately translates into an   increase   of    the  critical $t_\perp'^*$ and  an enhancement of d-wave Cooper pairing, as shown by the upward shift of the  $T_{c}$ line     in  Fig.~\ref{Phases}.

\subsection{Umklapp scattering}
\begin{figure}
\includegraphics[width=4.2cm]{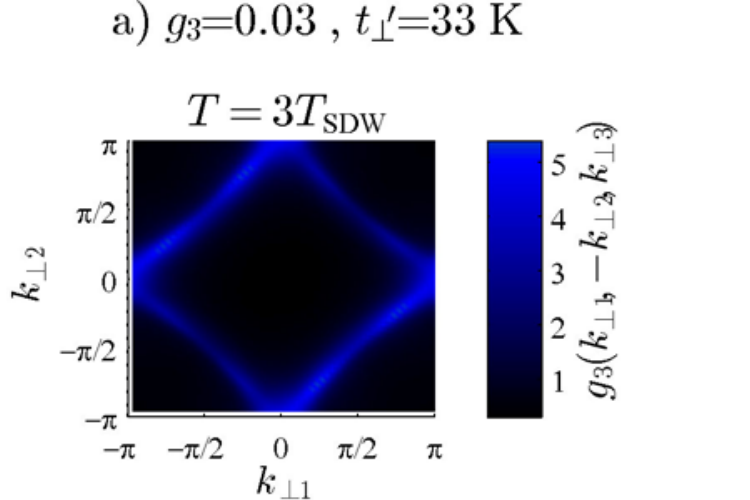}
 \includegraphics[width=4.2cm]{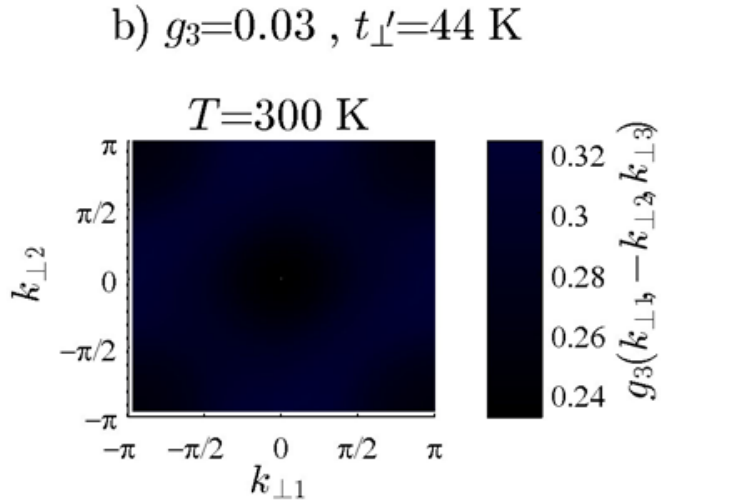}  \\ 
    \includegraphics[width=4.2cm]{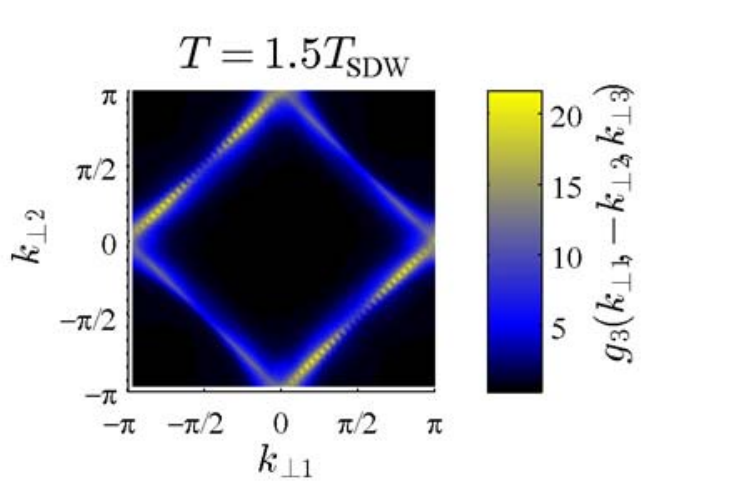}
     \includegraphics[width=4.2cm]{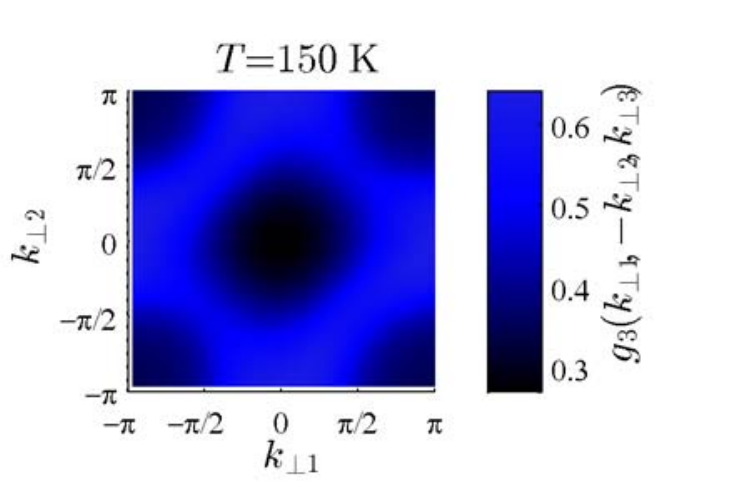} \\
      \includegraphics[width=4.2cm]{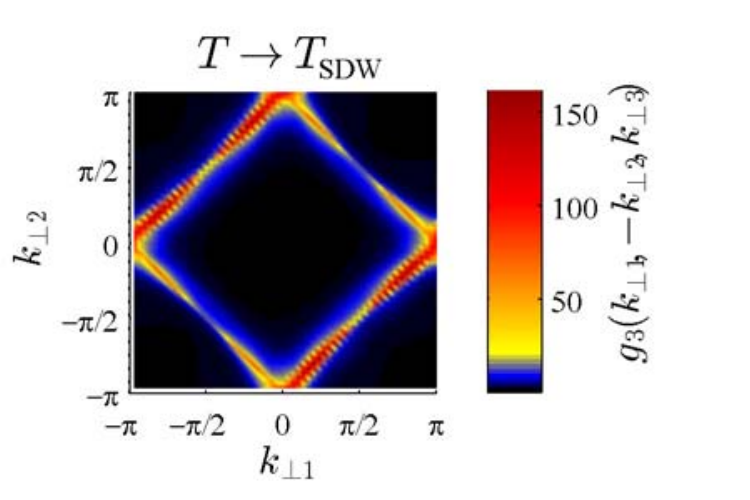}
     \includegraphics[width=4.2cm]{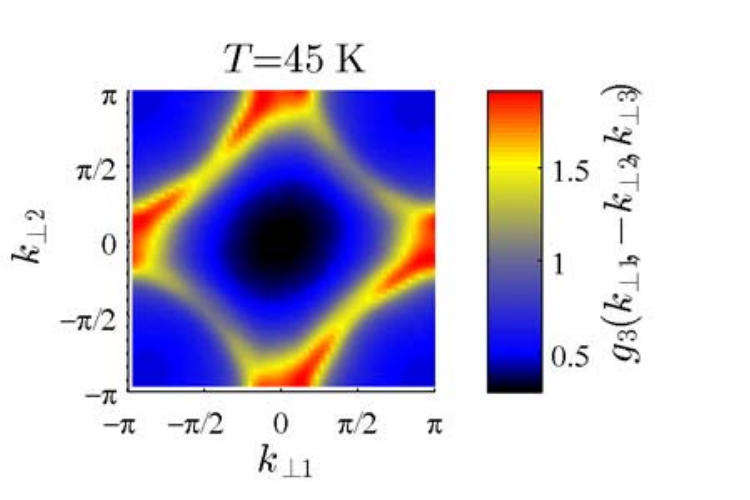} 
  \caption{Renormalized umklapp scattering amplitude projected in the $(k_{\perp1},k_{\perp2})$ plane  in  the normal phase at  different temperatures for the initial $g_3=0.03$ ($t_\perp'^* = 38$K). (a): $t_\perp'= 33$K $(<t_\perp'^*)$; (b) : $t_\perp'= 44$K $(>t_\perp'^*)$. \label{g3_1}}
\end{figure}

In the SDW sector of the phase diagram, the umklapp scattering amplitude $g_3$ develops like $\chi_{\rm SDW}$ a singular growth at the approach of   $T_{\rm SDW}$.  For $t_\perp' < t_\perp'^*$ and different temperatures, Fig.~\ref{g3_1}-a shows  the contour plot of $g_3(k_{\perp 1},-k_{\perp1},k_{\perp2})$ projected  in the $(k_{\perp1},k_{\perp2})$ plane at zero transverse momentum for the  pairs of ingoing and outgoing particles.  As shown in the Figure~\ref{g3_1}-a, the maximum of scattering intensity is mainly concentrated along the lines $k_{\perp2}= k_{\perp1} \pm \pi$, which is congruent with the  transverse  component of the SDW correlation  wave vector $\bm{q}_0=(2k_F,\pi)$.      Along these lines, the singular growth is the strongest around the spots $(0,\pm \pi)$ and $(\pm\pi,0)$, whose components   differ from the points $k_\perp=\pm\pi/4,\pm3\pi/4$ on the Fermi surface where  nesting is predicted to be optimal for the spectrum (\ref{Ek}). The fact that the warmest  regions of scattering   on the Fermi surface differ from the nesting prediction reflects the influence of Cooper scattering  channel in the anisotropic flow of $g_3$ in Eq.~(\ref{Flowg}).

\begin{figure}
\includegraphics[width=4.2cm]{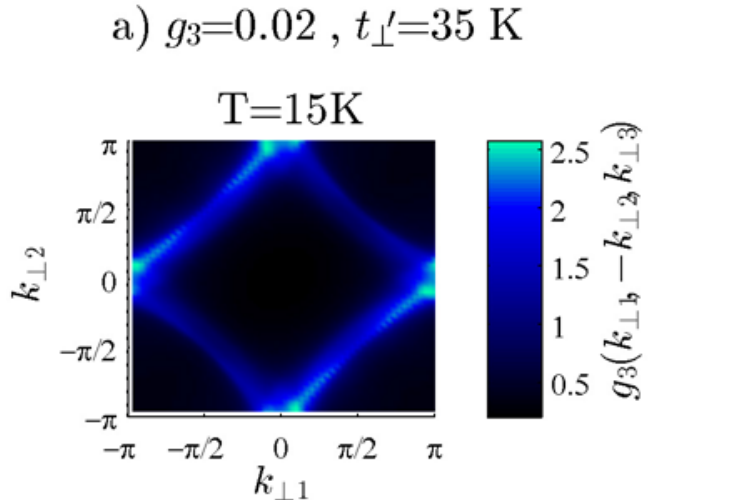}
 \includegraphics[width=4.2cm]{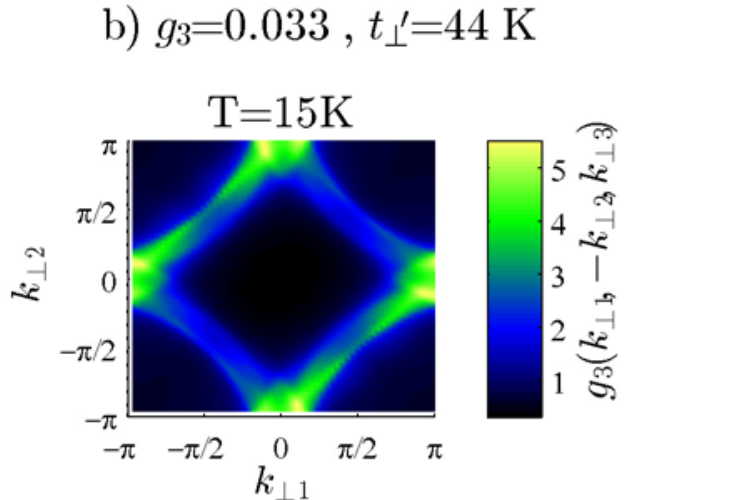}  \\ 
    \includegraphics[width=4.2cm]{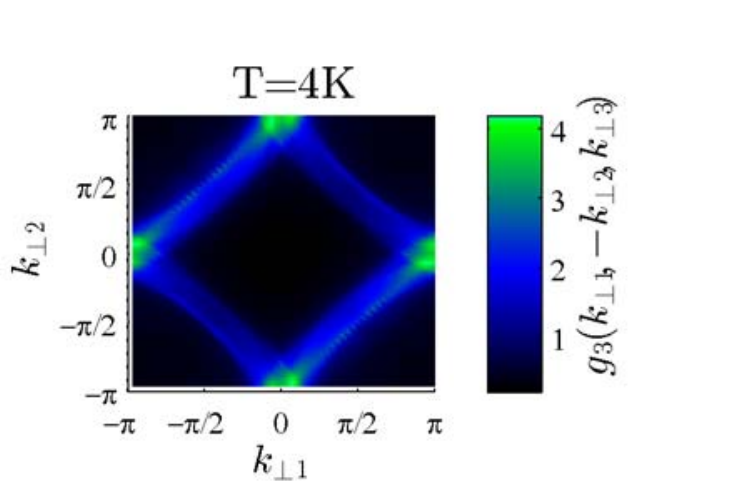}
     \includegraphics[width=4.2cm]{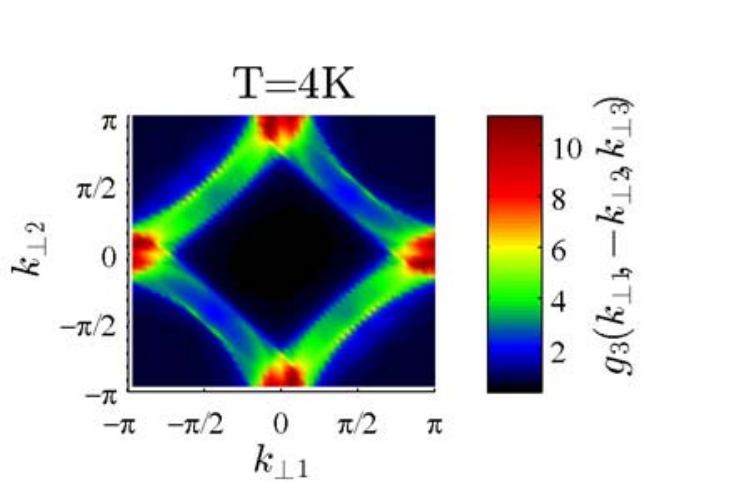} \\
      \includegraphics[width=4.2cm]{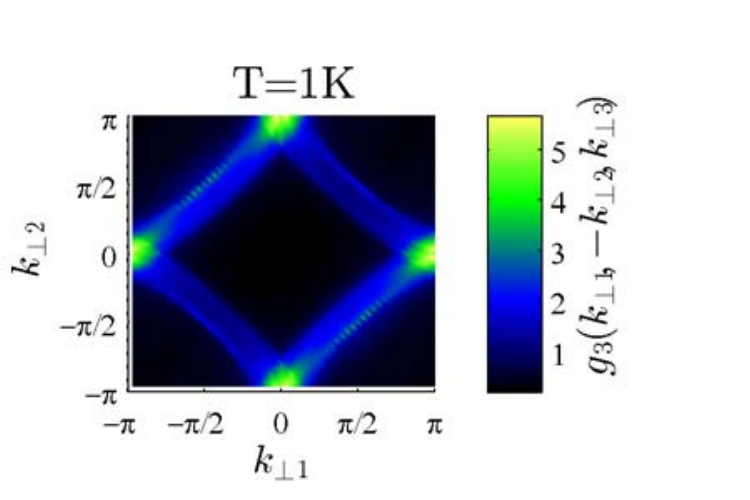}
     \includegraphics[width=4.2cm]{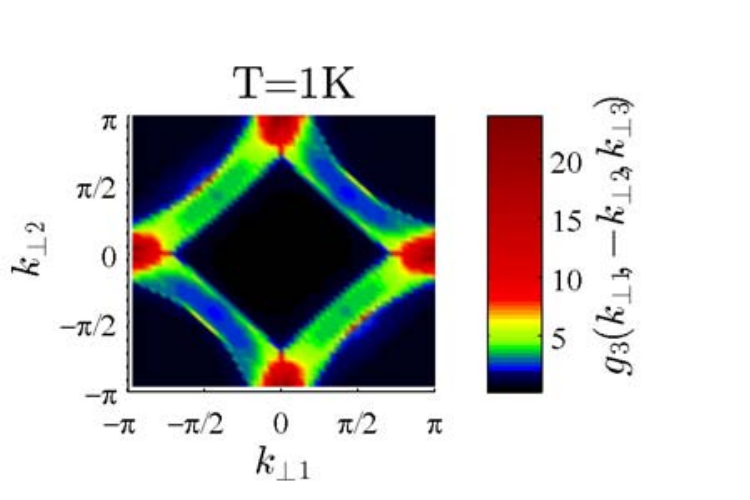} 
  \caption{ (Color online) Renormalized umklapp scattering amplitude projected in the $(k_{\perp1},k_{\perp2})$ plane  in  the normal phase of the SCd sector of the phase  diagram,as obtained  at  different temperatures and initial values of $g_3$.\label{g3SC}}
\end{figure}

If we now consider the low temperature variation of umklapp in the  SCd sector of the phase diagram at $t_\perp'>t_\perp'^*$, we note according to Fig.~\ref{g3SC} that $g_3$ is no longer singular for all temperatures down to $T_c$. However, as temperature is decreasing we still observe an enhancement of $g_3$ along the SDW lines of scattering  $k_{\perp2}= k_{\perp1} \pm \pi$. These are associated with peaks at $k_\perp=0,\pm \pi$ on the Fermi surface whose amplitude  increases   with the bare   $g_3$, as shown in Fig.~\ref{g3SC}. According to the combination $g_{\rm SDW}= g_2+ g_3$ of couplings entering in the SDW susceptibility [see Eq.~(\ref{gmu})], a positive increase  in umklapp goes hand in hand with  an increase  of SDW spin correlations.
This has been shown  to yield a  temperature dependence for $\chi_{\rm SDW}(\bm{q}_0)\sim 1/(T+ \Theta)$ of the Curie-Weiss form  in the same temperature domain, with an energy scale $\Theta$ for SDW correlations approaching zero at $t_\perp'^*$ and raising linearly with $t_\perp'$ from the QCP \cite{Bourbon09,Sedeki12,Bourbon11}. Occurring despite poor nesting conditions, the enhancement is the consequence of a positive feedback of d-wave Cooper pairing on SDW correlations. As mentioned earlier, this is made possible by a $k_\perp$-dependent coupling of $g_3$ to normal scattering processes $g_2$ and $g_1$. According to (\ref{Flowg}), the first Fourier component in $\cos (k_{\perp1}d_\perp)\cos (k_{\perp2}d_\perp)$ of these couplings grows as temperature is lowered to finally become   singular   at the approach of $T_c$  where $ \chi_{\rm SCd}(0)$ is divergent. 

Anisotropy of  umklapp scattering in  momentum space along with its temperature dependence enter as  key factors in the determination of   resistivity at low temperature.

\section{Numerical results for the Boltzmann equation}
We can now proceed to   the numerical solution of the linearized semi-classical Boltzmann equation (\ref{LinBE}), using (\ref{LBEb})  and the above RG results for the renormalized umklapp vertex part. The solutions for the normalized functions $\bar{\phi}_{k_\perp}$  have been obtained for $N_P=60$ patches for the Fermi surface sheet $\bm{k}_F^p(k_\perp)$, which assures a convergence for the solutions of the Boltzmann equation (\ref{LinBE}) or (\ref{LBEb}) that is independent of the discretization of the transverse Brillouin zone.  Following (\ref{sigma}),  the temperature dependence of longitudinal resistivity can then be obtained by tuning the values of $t_\perp'$ and initial $g_3$. 

\subsection{Dimensionality of the electron gas and resistivity}

\begin{figure}
 \includegraphics[width=3in]{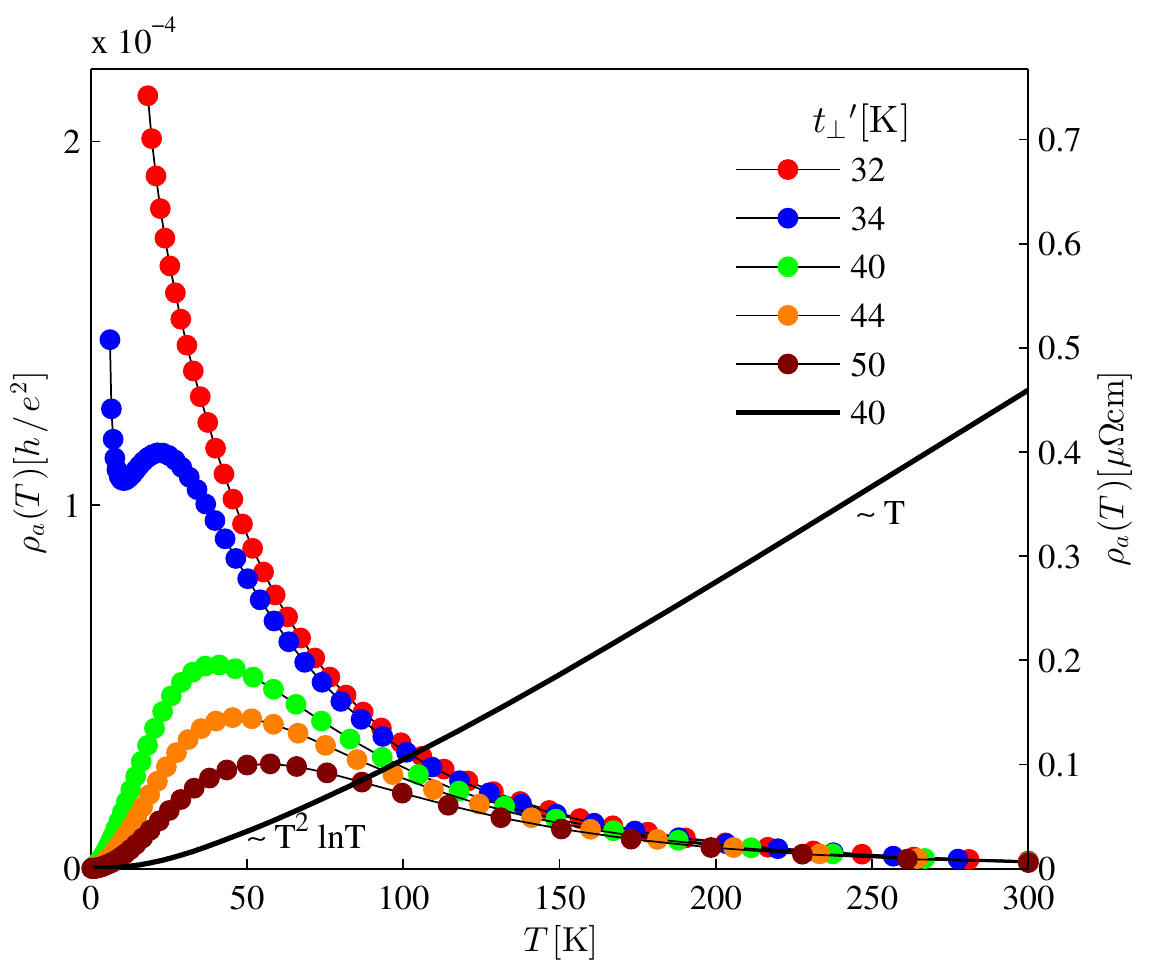}
\caption{(Color online) Calculated  longitudinal resistivity  over the whole  temperature interval for different $t_\perp'$ at $g_3=0.03$ ($t_\perp'^* = 38$K). The continuous line refers to the Fermi liquid limit at constant scale-independent $g_3(= 0.03)$ yielding linear $T$-resistivity in the high temperature 1D domain $(T>t_\perp)$ followed by a crossover toward a 2D $T^2\ln T$ regime at low temperature. }
 \label{RhoaFT}
 \end{figure}

  We first consider in Fig.~\ref{RhoaFT} the general features of longitudinal resistivity over the whole temperature interval. This  includes the high - effectively 1D - temperature domain at $T>t_\perp$, its crossover to  the 2D region  at $T< t_\perp$  where the warping of the Fermi surface becomes quantum mechanically coherent, and finally   the temperature domain  $T_\mu<T<t_\perp'$ where antinesting effects come into play at the approach of instabilities.
  
   From the Figure~\ref{RhoaFT}, we see  that in the 1D temperature region of both SDW and SCd sectors of the phase diagram, the resistivity is not metallic but grows with decreasing temperature. This is the result of  relevant   umklapp scattering whose flow in the 1D domain is characterized  by essentially no $k_\perp$ dependence, that is,   no transverse correlations, as illustrated in Fig.~\ref{g3_1}-b (top panel). The growth  
completely overcomes   the linear $T$-resistivity decrease obtained  in the limit of a constant $g_3$ (continuous line of Fig.~\ref{RhoaFT}). This insulating behaviour is consistent with previous 1D results obtained  at half-filling \cite{Gorkov73,Giamarchi91}. 
  
When $t_\perp'< T < t_\perp$, the system crossovers to its 2D regime with   antinesting effects remaining weak. In this temperature domain, the relevance of $g_3$ is still strong enough to outweigh  the crossover toward the low temperature Fermi liquid $T^2\ln T$-resistivity behaviour   found at constant $g_3$\cite{Gorkov98,Zheleznyak99} (continuous line of Fig.~\ref{RhoaFT}). Resistivity remains insulating in this region.
  It is only when $T$ is further lowered below $t_\perp'$ that the anisotropic growth of $g_3$ is significantly damped.  At $ t_\perp'>t_\perp'^*$, in the SCd sector where umklapp, though still increasing, is no longer singular at finite $T$,  resistivity goes through a maximum at $T\sim t_\perp'$ and is followed at lower temperature by   a metallic   behaviour.  Metallic behaviour is also possible above $T_{\rm SDW}$ in the SDW  sector of the phase diagram  provided $t_\perp'$ is relatively close to the critical $t_\perp'^*$,  as portrayed  in Fig.~\ref{RhoaFT}. As we will show next, however, metallicity is in general not equivalent to a Fermi liquid behaviour. Deviations from the latter are found and correlated to the distance from $t_\perp'^*$;  it is only at strong $t_\perp' \gg t_\perp'^*$ that Fermi liquid features  tend to be restored at sufficiently low temperature.

\subsection{Resistivity at quantum criticality and beyond}
\begin{figure}  
\includegraphics[width=8cm]{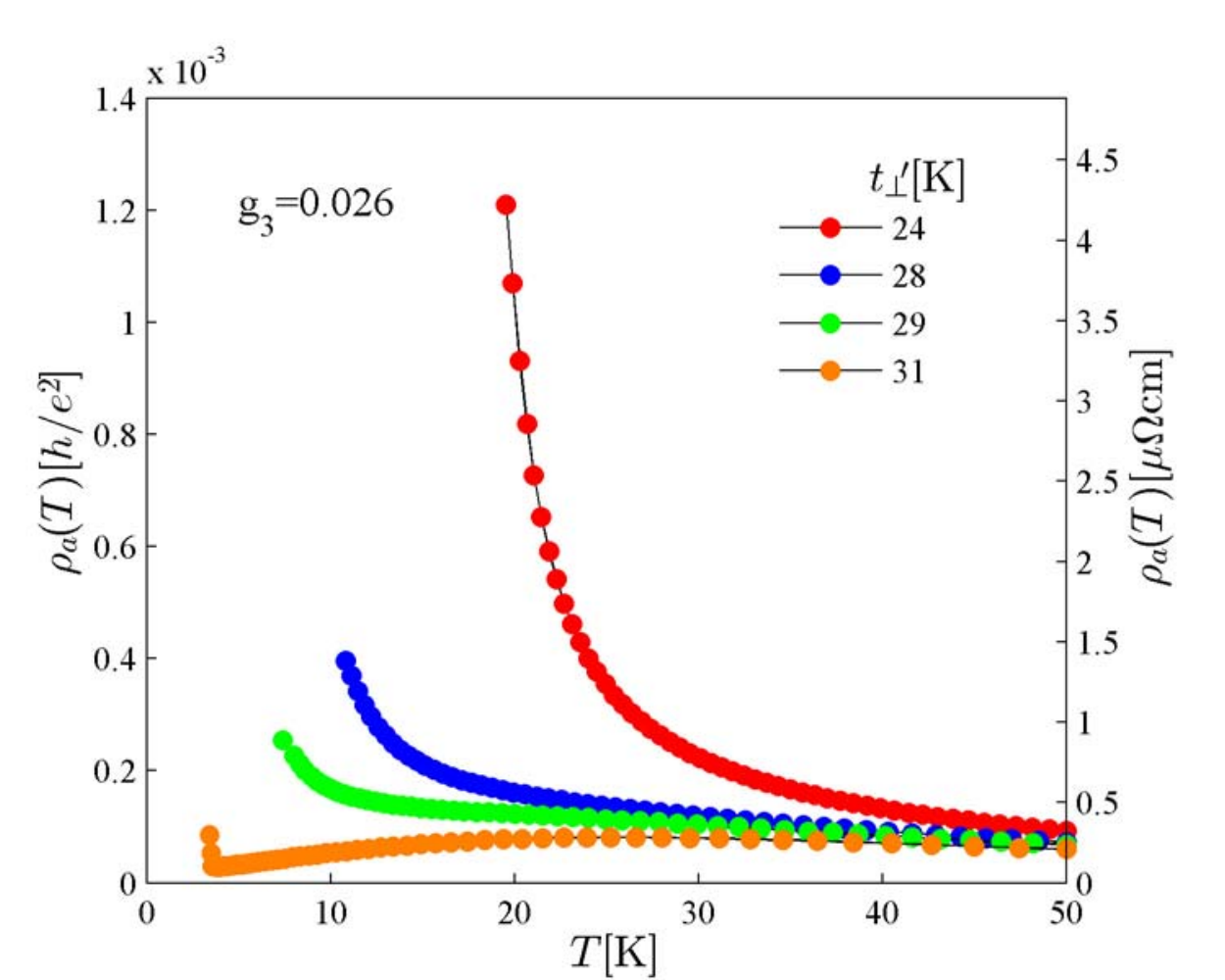} 
\includegraphics[width=8cm]{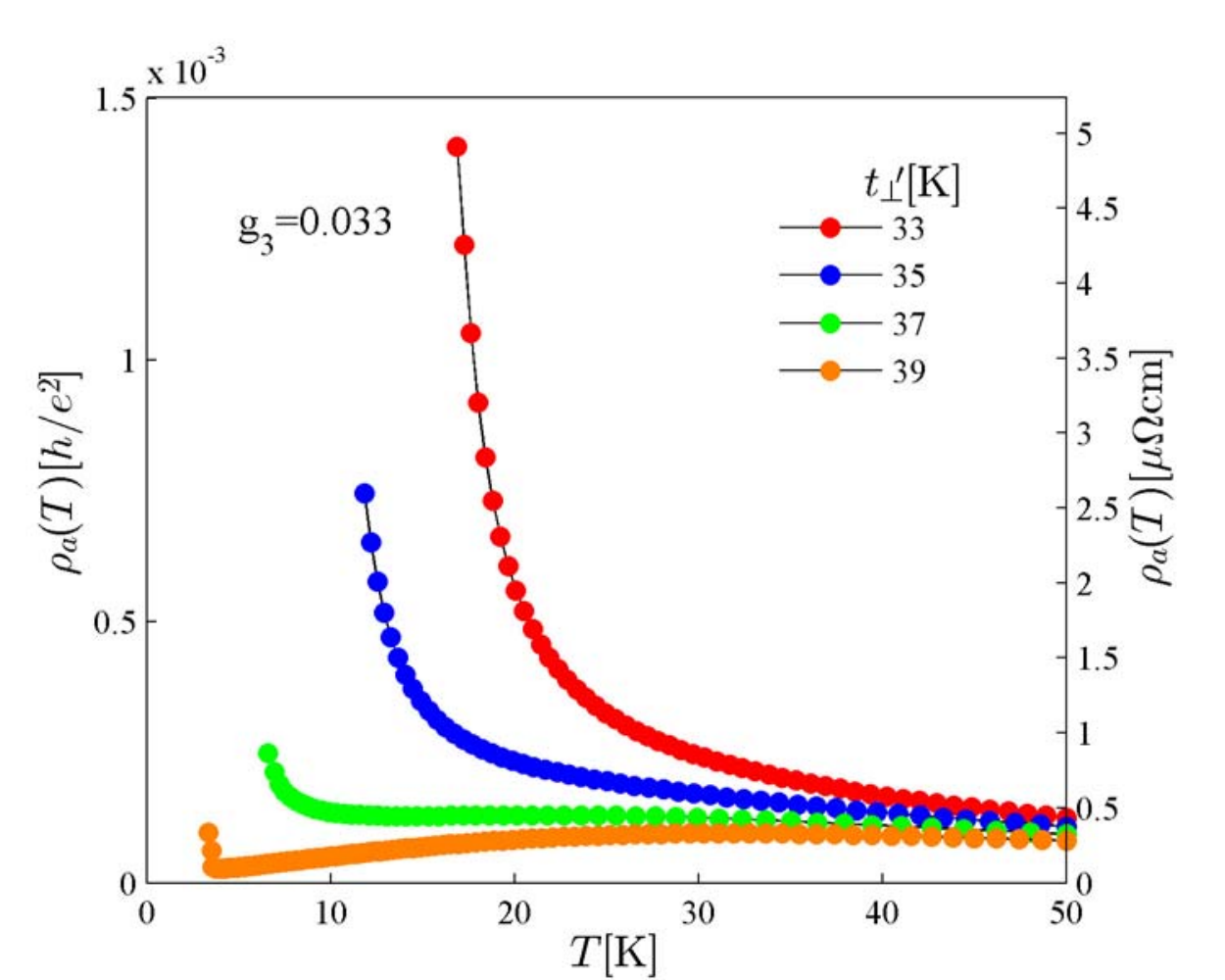}
\caption{(Color in line) Calculated resistivity at low temperature in the SDW domain of the phase diagram for different  $t_\perp'$ and $g_3$.}
\label{RhoSDW}
\end{figure}

We now focus on the  low temperature profile of $\rho_a$ at $T_\mu < T< t'_\perp$  on either side of the QCP  $t_\perp'^*$. Starting with the SDW side of the phase diagram, the Fig.~\ref{RhoSDW} displays the temperature dependence of resistivity in the normal phase of the SDW instability at different $t_\perp^\prime$ and bare $g_3$ values. When $t_\perp'$ is relatively  far below $t_\perp'^*$, resistivity remains insulating like with $d\rho_a/dT <0$ down to $T_{\rm SDW}$. The growth of umklapp scattering and  therefore of SDW fluctuations is sufficiently  strong to maintain the source of inelastic scattering and the  insulating response of the electron gas. In these conditions, the behaviour varies weakly with the initial amplitude of $g_3$.

\begin{figure}  
\includegraphics[width=8cm]{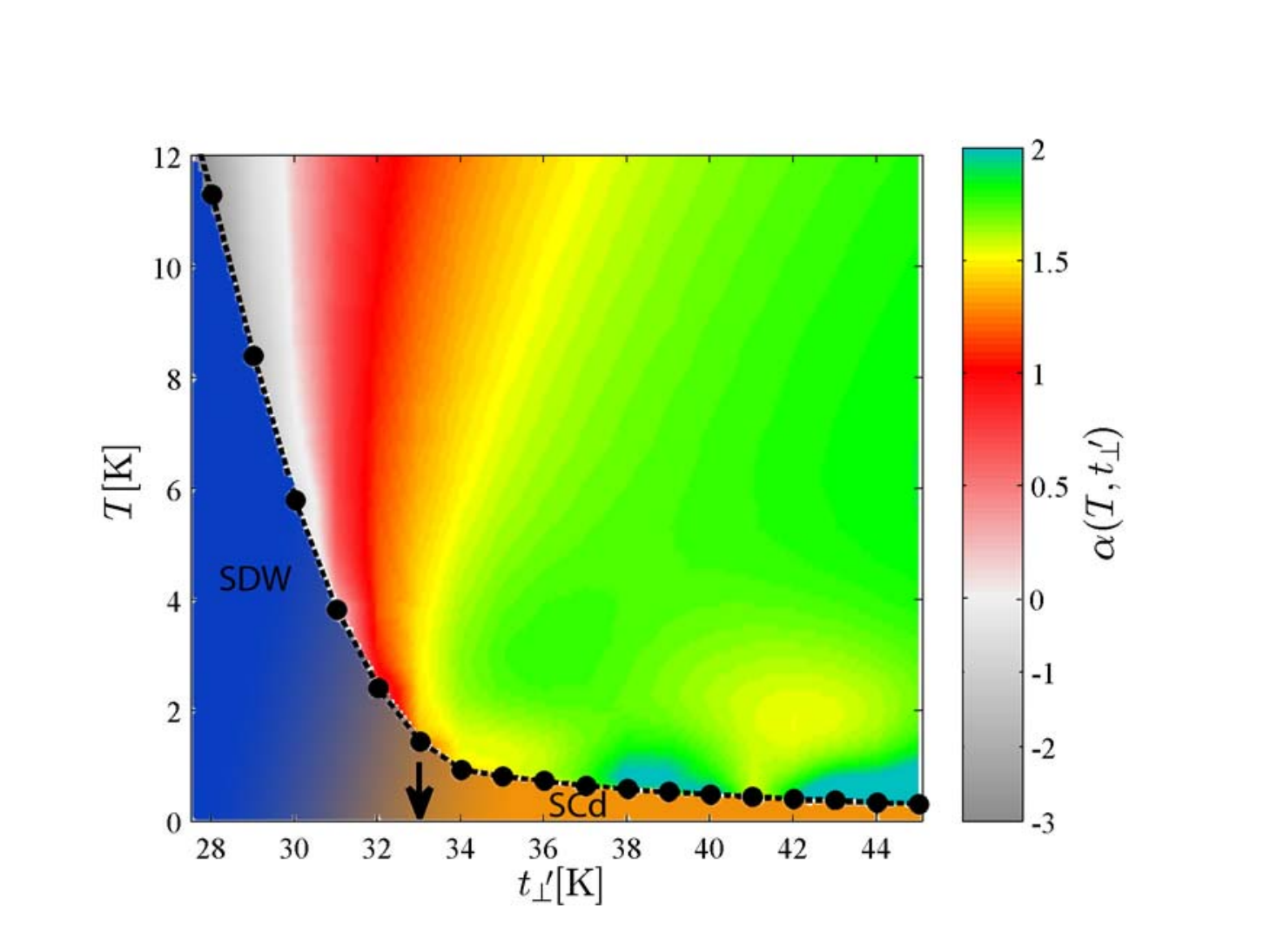} \vskip 0.5cm 
\includegraphics[width=8cm]{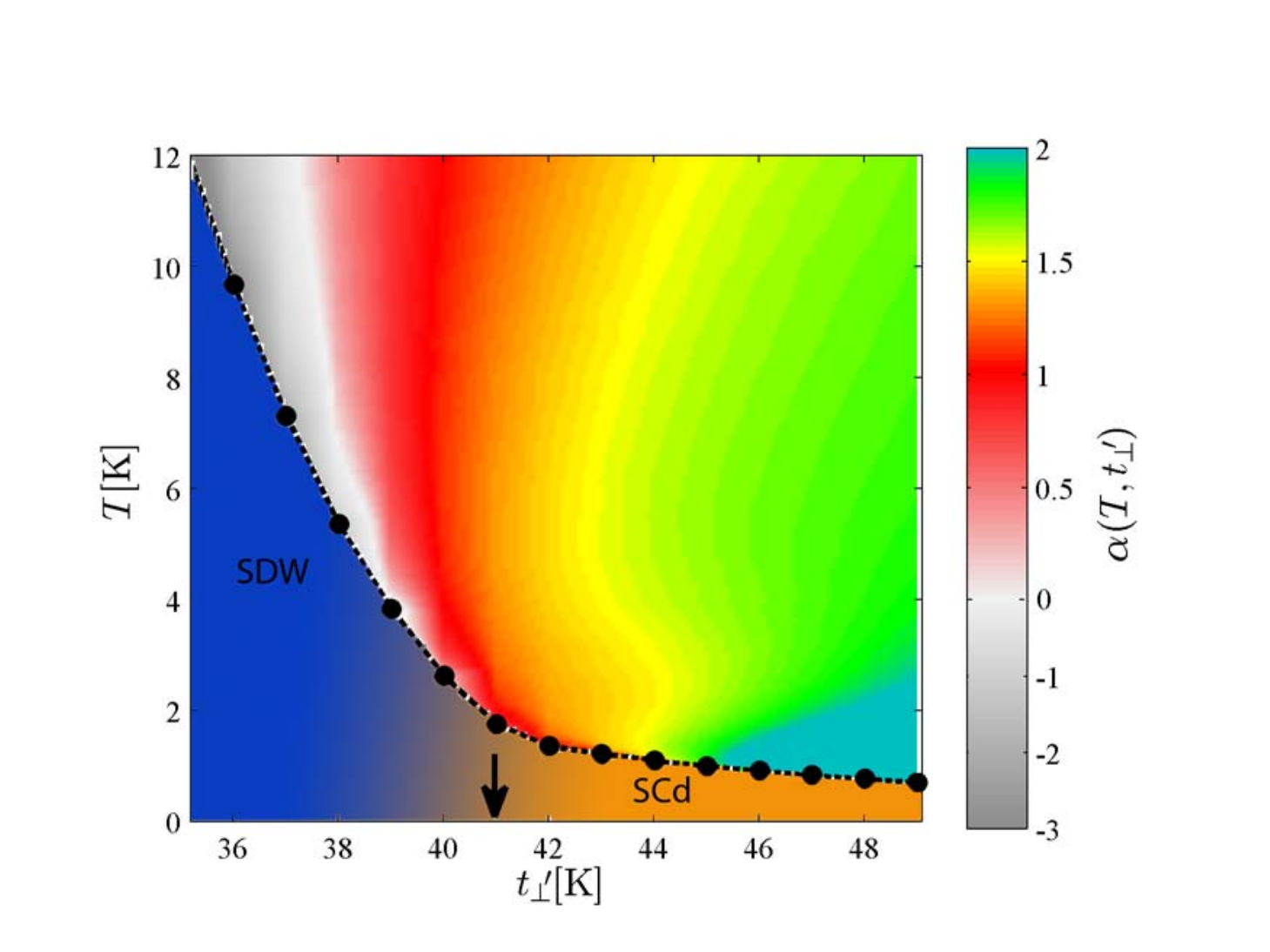}
\caption{(Color online) Variation of the power law exponent $\alpha$ of resistivity in the phase diagram for $g_3=0.026$ (top) and $g_3=0.033$ (bottom). The arrows indicate the location of $t_\perp'^*$.}
\label{Alpha_T}
\end{figure}

 The temperature profile of resistivity qualitatively  change at the approach of $t_\perp'^*$. At $T\sim t_\perp'$,  $\rho_a$ reaches a maximum followed by a crossover to a metallic behaviour which persists down to very close to $T_{\rm SDW}$.  The singularity in umklapp, being still present despite sizeable nesting alterations,     generates a resistivity upturn due to  critical scattering, as meant in Figs~\ref{RhoaFT} and \ref{RhoSDW}. In the  metallic regime, however, the resistivity deviates from the Fermi liquid prediction. A suitable way to describe these deviations is to express resistivity  as  a power law, $\rho_a(T)\sim T^{\alpha(T)}$, with a local exponent $\alpha(T)$ defined by   
\begin{equation}
\label{alpha}
\alpha(T) = {d\over d\ln T}\ln\rho_a(T).
\end{equation}
The values of the exponent $\alpha$ are inserted  in the calculated phase diagrams and shown in Fig.~\ref{Alpha_T} for different bare amplitudes of $g_3$. Thus in approaching  $t_\perp'^*$ from below the exponent $\alpha(T)$ crossovers from negative (insulating)  to positive (metallic) values. Sufficiently close to $t_\perp'^*$,  $T_{\rm SDW}$ is small and $\alpha(T)$ reaches unity for $T\lesssim 10$K, which corresponds to the linear-$T$ resistivity usually associated to the QCP.  

This linearity can be qualitatively understood if one considers that at low temperature resistivity is roughly speaking proportional to  $T^2\langle g^2_3\rangle_{\rm FS} $. This would indicate that   the average of the square of  umklapp scattering over the Fermi surface, $\langle g^2_3\rangle_{\rm FS}\sim 1/T$, essentially grows as the inverse of  temperature above the transition.  Note that this temperature dependence is similar to the SDW susceptibility  that goes  like $\chi_{\rm SDW}\sim 1/T$ at $T\gg |\Theta|$ in this region \cite{Bourbon09,Sedeki12}.

\begin{figure*}  
\includegraphics[width=8cm]{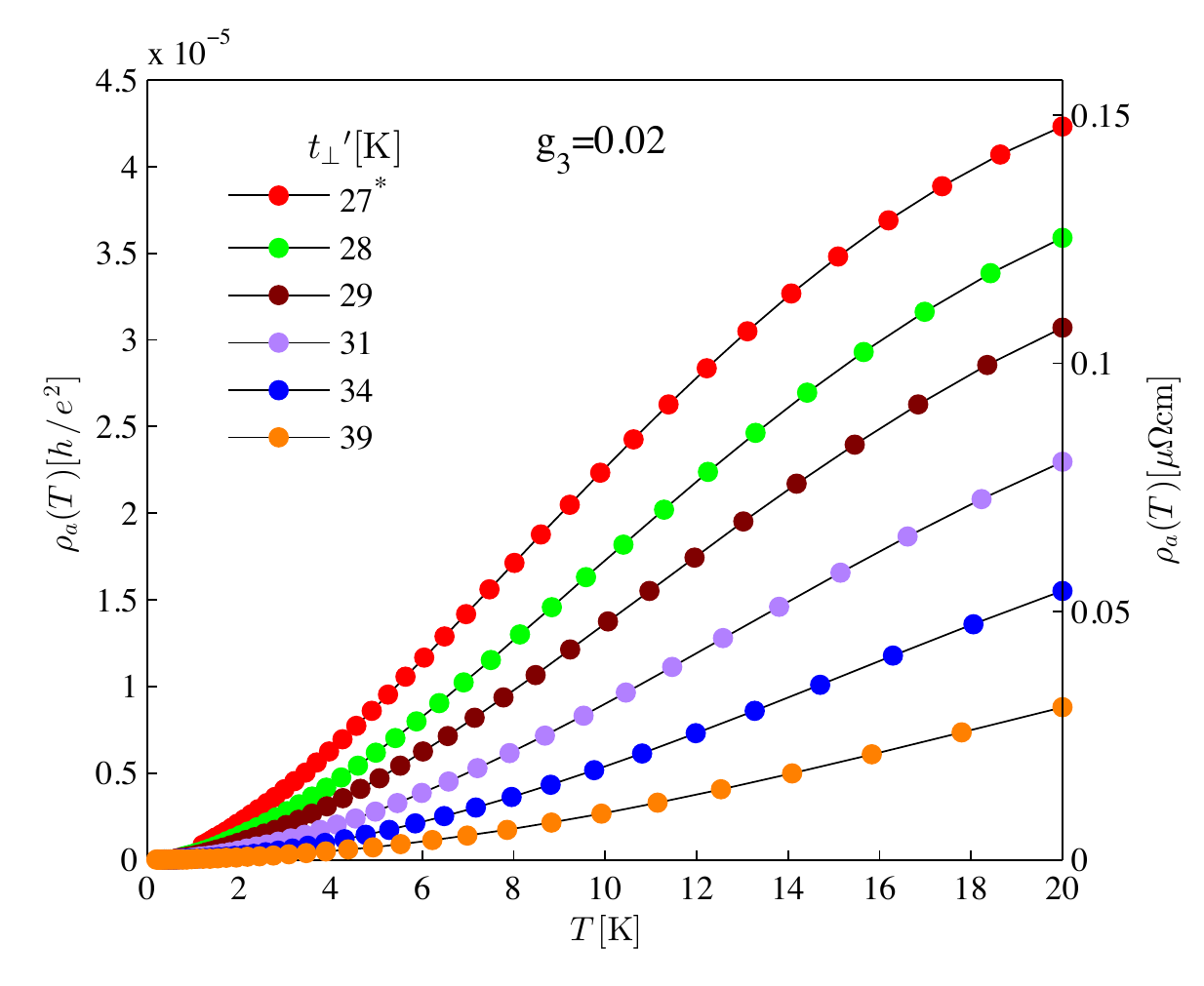} \hfil
\includegraphics[width=8cm]{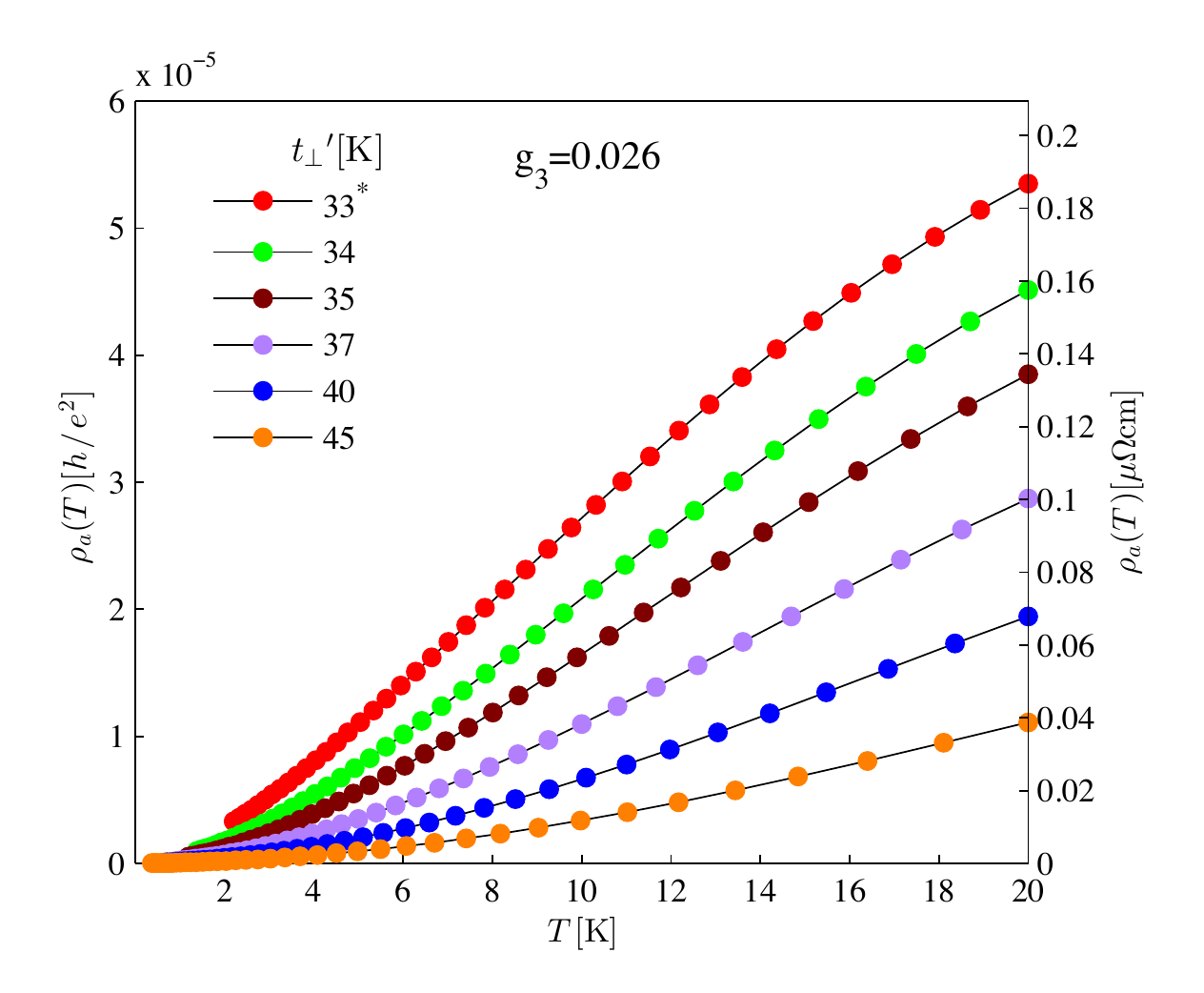}\\
\includegraphics[width=8cm]{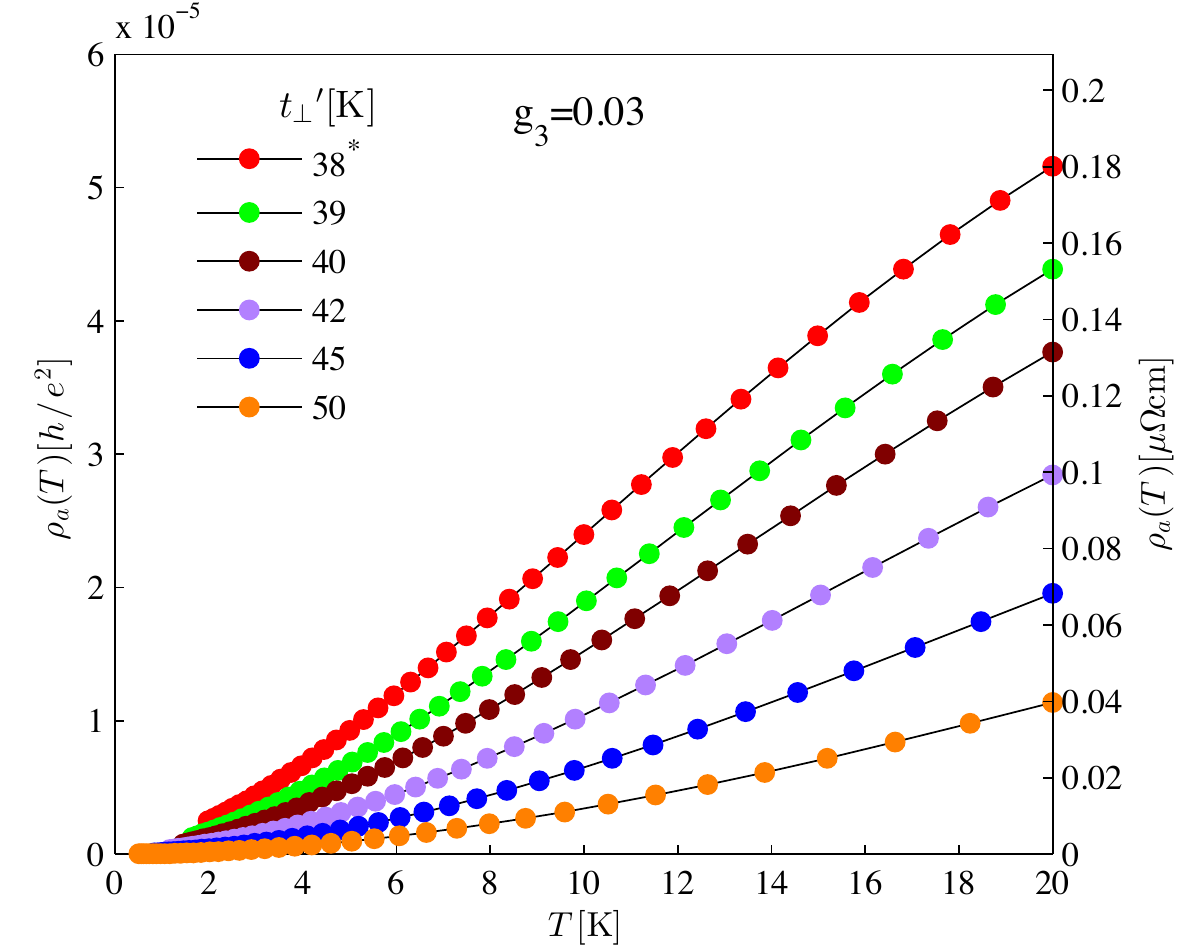} \hfil
\includegraphics[width=8cm]{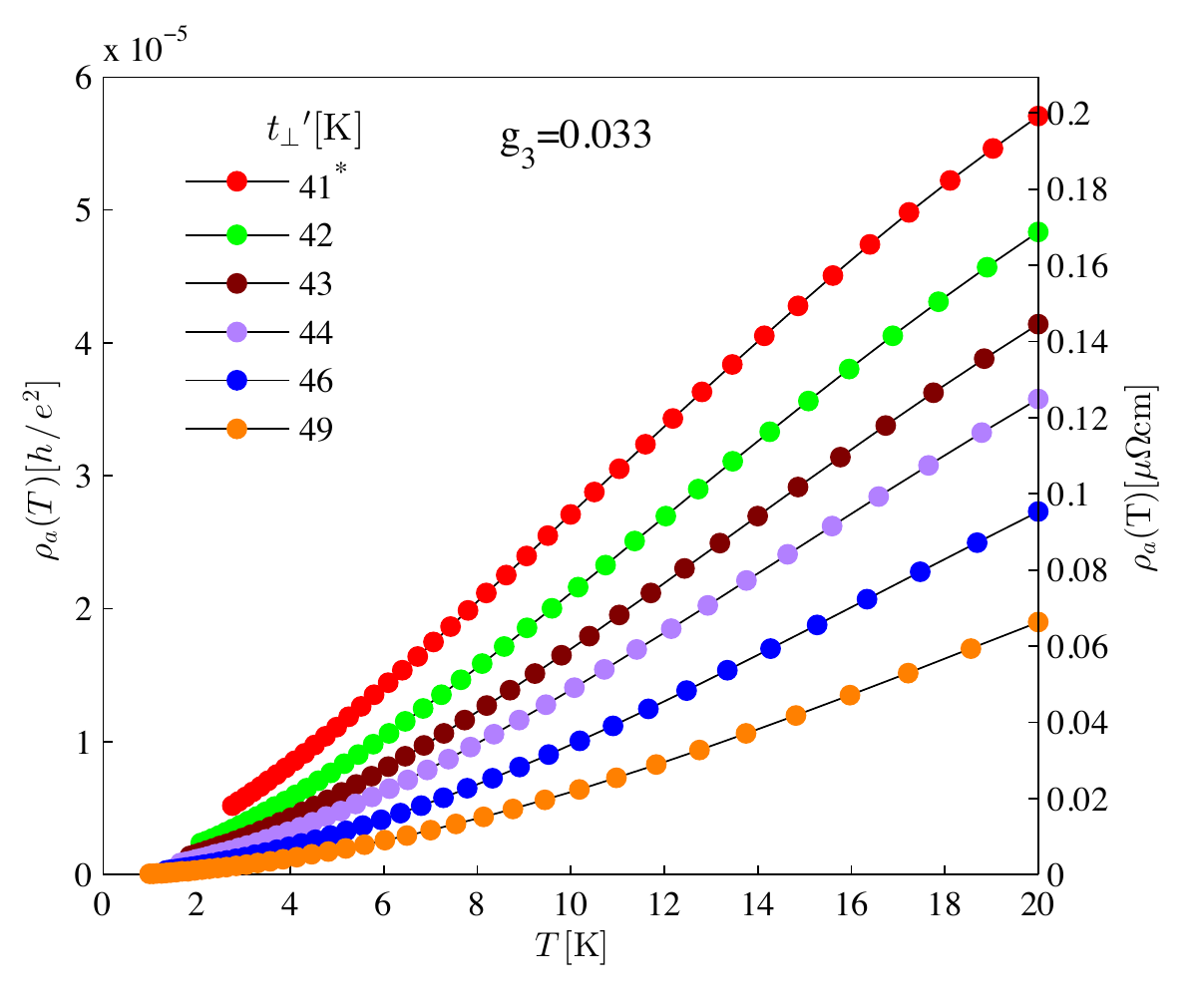}
\caption{ (Color online) Calculated resistivity at low temperature in the SCd domain of the phase diagram for different  $t_\perp'$ and $g_3$.}
\label{RhoSCd}
\end{figure*}

If we now consider  the SCd region at $t_\perp'\ge t_\perp'^*$, the linearity continues to be seen over a finite region of the phase diagram. From Figures \ref{RhoSCd} and \ref{Alpha_T}, it can be found over a large temperature interval above $T_c$   near $t_\perp'^*$. However, a shift  in the value of $\alpha(T)$ toward  larger values  takes place at     low enough  temperature especially if $g_3$ is relatively small initially. As a function of temperature a fan shape like region  then unfolds from the neighbourhood of $t_\perp'^*$ in which $\alpha(T)\simeq 1$. This region of linear resistivity in the phase diagram broadens as the bare $g_3$ increases  and spin fluctuations in the normal phase and then $T_c$ are enhanced.  Thus contrary to the situation expected for a classical   QCP tied to SDW ordering alone\cite{Moriya03}, this region of linearity is not quickly followed as a function of $t_\perp'$ by  a $T^2$-resistivity. The   crossover region  $1<\alpha(T)<2$ where  deviations  are present is rather wide and covers the whole region where superconductivity is present, as portrayed in  Figure~\ref{Alpha_T}  for different $g_3$. It is only in the low temperature domain close to $T_c$ and for $t_\perp'$ well above   $t_\perp'^*$   that the boundary with the Fermi liquid  limit appears and  $T^2$-resistivity  becomes visible (see Figures \ref{RhoSCd} and \ref{Alpha_T}), namely where $\langle g_3^2\rangle_{\rm FS}$ becomes essentially temperature independent. Similar features were found in previous RG calculations  of the electron-electron scattering rate that included  both normal and umklapp processes\cite{Sedeki12}. These were attributed to the presence of the superconducting instability that still influence the growth of the coupling constants; it gives rise to a  positive feedback of SCd pairing on umklapp scattering which can account for this extended quantum critical behaviour.    
\subsection{Anisotropy of the scattering rate}
It is instructive to extract from the above results  for resistivity some information about the momentum resolved - $k_\perp$-dependent - scattering rate. According to (\ref{sigma}), the conductivity  $\sigma_a=\sum_i \sigma_{k_{\perp i}}$ is a sum of   contributions coming from each patch. One can associate to each contribution a patch resistivity $ \rho_{k_{\perp i}} = (\sigma_{k_{\perp i}})^{-1}$ as  an element  of an array of parallel resistors   proportional to a scattering rate at $\bm{k}_F^p(k_{\perp i})$ on the Fermi surface.  

\begin{figure}  
\includegraphics[width=9cm]{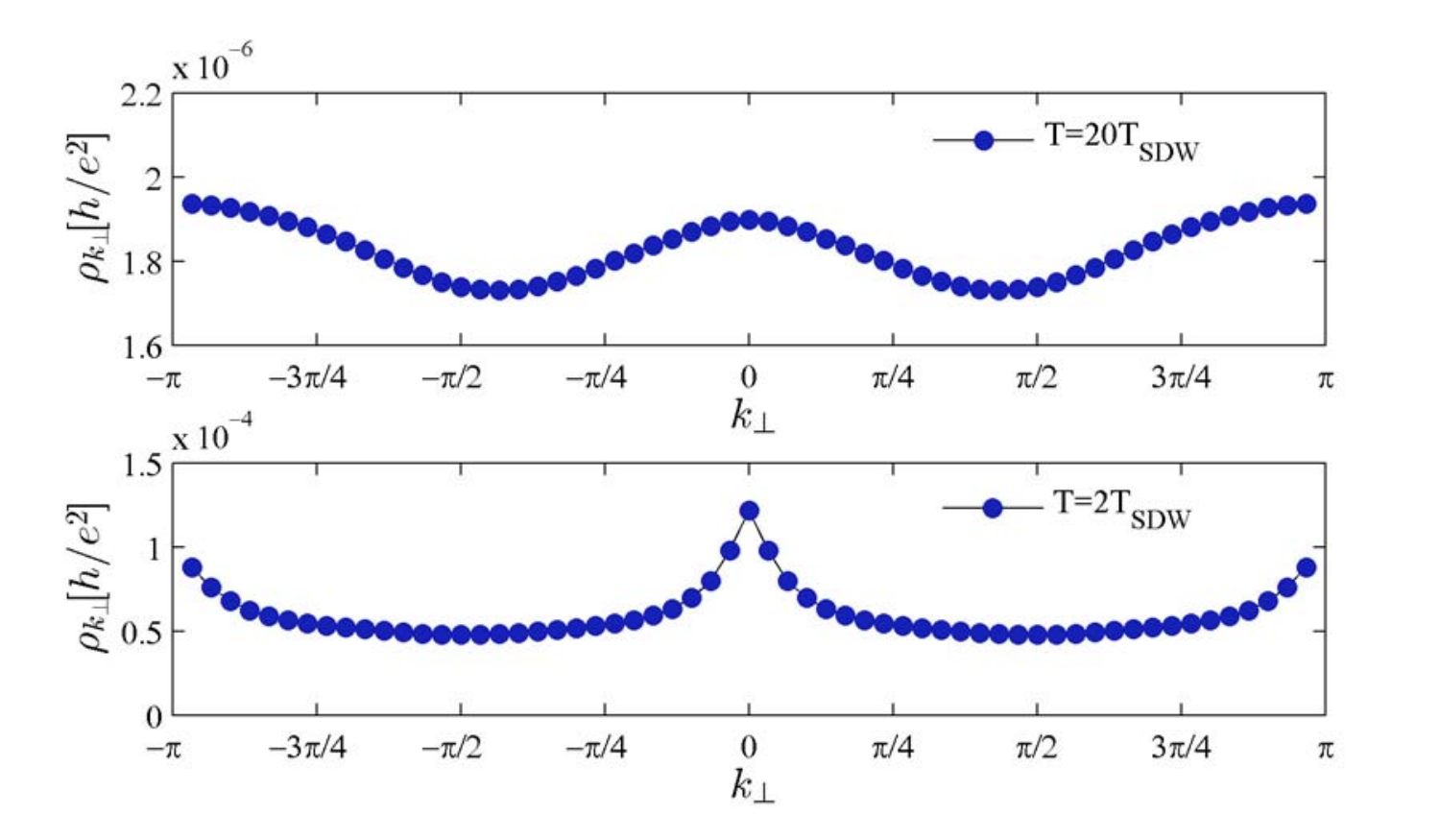} \\
\includegraphics[width=9cm]{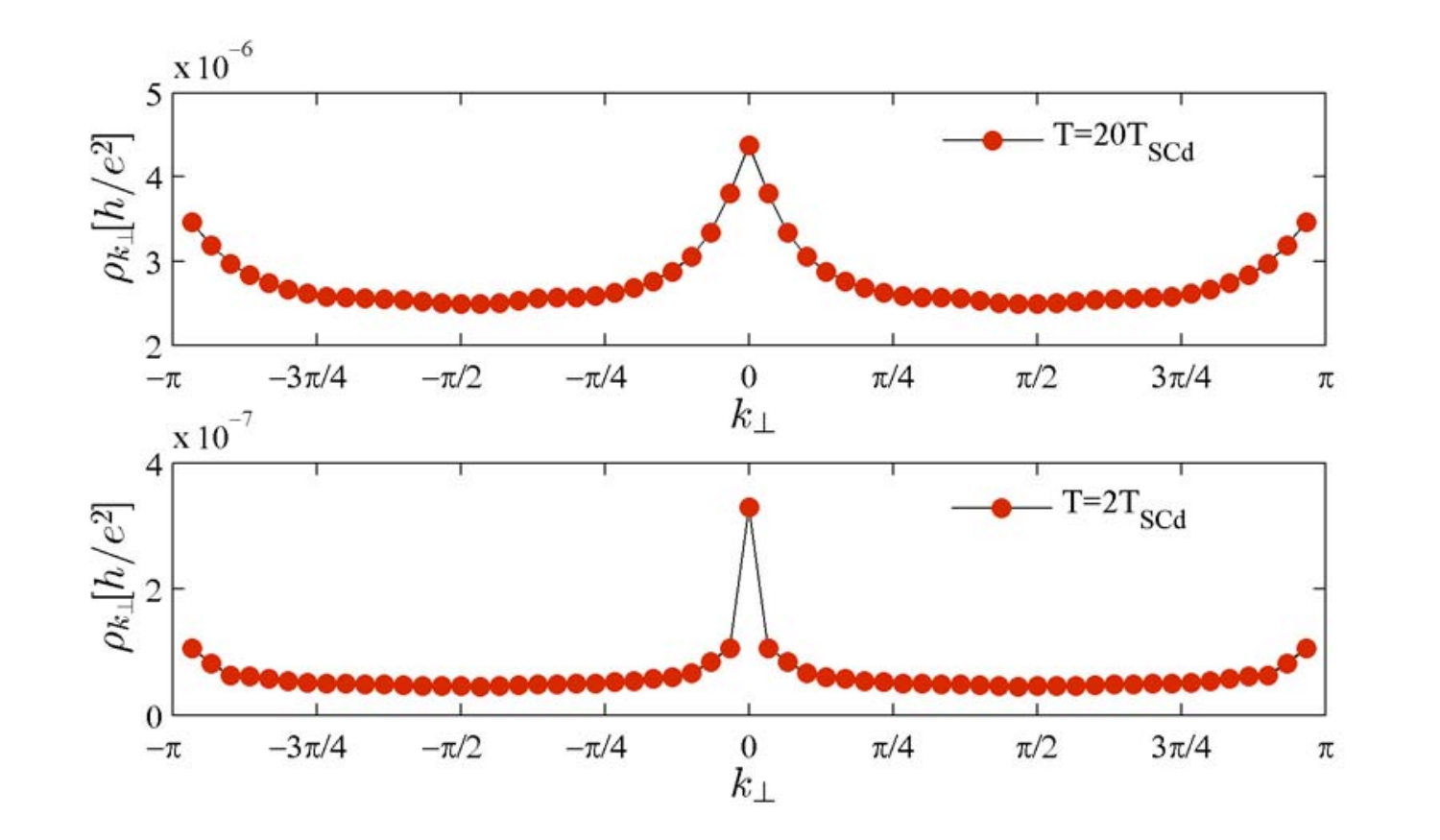}
\caption{(Color online) Variation of resistivity components along the Fermi surface at different temperatures: SDW  (top) and  SCd (bottom) domains.\\}
\label{Anis}
\end{figure}

The Figure~\ref{Anis} shows the variation of $\rho_{k_{\perp i}}$ as a function of $k_{\perp i}$. In the SCd regime, there are  peaks at $k_\perp=0$ and $\pm \pi$ indicating that the scattering rate is larger at these points. These warmer regions of scattering are the same as those  of umklapp scattering amplitude along the Fermi surface in Fig.~\ref{g3SC} and then differ from the positions $k_\perp = \pm \pi/4$ and $\pm3\pi/4$  attributed to the best nesting conditions for the spectrum. The resulting momentum profile of $  \rho_{k_{\perp i}}$ over the Fermi surface is also congruent with the one previously found by a direct RG calculation of the scattering rate from the one-particle self-energy at low temperature, which takes into account both normal and umklapp processes \cite{Sedeki12}. As stressed in Ref.~\cite{Sedeki12} these warmer regions of scattering turn out to coincide with the locations of the maximum for a d-wave gap  along the Fermi surface.  Since a single   scattering channel (Peierls) calculation would invariably leads to the best nesting points as the warmest spots of scattering on the Fermi surface, it follows that one can ascribe the shift in the spots on the Fermi surface to the presence of the interfering Cooper channel in the complete one-loop calculation (\ref{Flowg}).  The interference between both scattering channels being present in the SDW sector of the phase diagram as well, the  anisotropy of $\rho_{k_\perp}$ is  found to be similar  above $T_{\rm SDW}$, as displayed in Fig.~\ref{Anis}.

\subsection{Comparison with experiments}

\begin{figure} 
 \includegraphics[width=9cm]{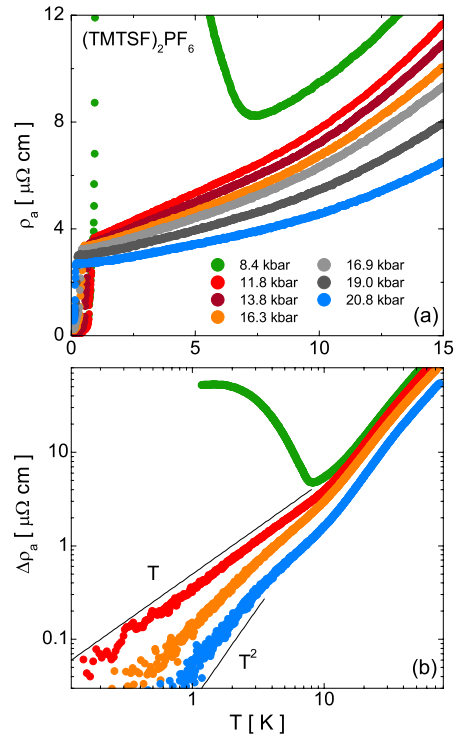}
 \caption{(Color online) (a): Longitudinal resistivity of Ref.~\cite{DoironLeyraud09} observed at different pressures or $T_c$ for (TMTSF)$_2$PF$_6$; (b): Inelastic part of the data ($\Delta \rho_a=\rho_a-\rho^0_a$) traced in a log-log plot showing the growth of the power law exponent $\alpha$ for resistivity with pressure. \label{rhoaPF6} }
 \end{figure}

In the following we shall make some comparison of the above results with   experiments. Starting with the SDW sector of the calculated phase diagram at low nesting deviations, the  results displayed  in Figures~\ref{RhoaFT} and \ref{RhoSDW} show the absence of metallic resistivity above $T_{\rm SDW}$. This is  at variance to what is actually observed  in system  like (TMTSF)$_2$PF$_6$ series of the Bechgaard salts series  below $P_c$ where a nearly $T^2$ behaviour for resistivity is well known to be observed above the resistivity upturn at $T_{\rm SDW}\sim10$K\cite{Bechgaard80,Moser98,Vuletic02,DoironLeyraud09} (see also Fig.~\ref{rhoaPF6}). The temperature profile obtained for resistivity well below $t_\perp'^*$ is rather characteristic of the one seen in the cousin sulfur based compounds (TMTTF)$_2$X (X=PF$_6$, Br, ...) at low pressure\cite{Coulon82}, which are characterized by a  1D Mott insulating gap at high temperature due to umklapp scattering, followed at lower temperature by a transition to an antiferromagnetic ground state\cite{Balicas94,Klemme95,Moser98}. Under moderate pressure, the normal phase of these systems becomes metallic with a power law behaviour down to $T_{\rm SDW}$ which would agree qualitatively  with  the calculated  positive values for  $\alpha$ found at low temperature not too far below $t_\perp'^*$ in  Figure~\ref{Alpha_T}. However in contrast to the results shown in Fig.~\ref{RhoSDW}, no maximum in resistivity is found experimentally  for longitudinal resistivity at a scale that would correspond to $t_\perp'$.\cite{Moser98}. These discrepancies provide some indication that inelastic scattering   at intermediate temperature and above would come  from processes  other than half-filling  umklapp\cite{Giamarchi97,Schwartz98,Tsuchiizu01}. 

On the SC side, above the QCP at $P_c$, detailed resistivity measurements for (TMTSF)$_2$PF$_6$ have been performed in the   low temperature domain up to about 15 times $T_c  $ ($T_c$ of the order of 1~K at $P_c$)  \cite{DoironLeyraud09,DoironLeyraud10}.   The resistivity data are reproduced in Figure~\ref{rhoaPF6}. At the pressure of 11.8 kbar, close to the $P_c\simeq 9$kbar, resistivity is found to be $T-$linear ($\alpha =1$) from  up to 10$T_c$ down to $T_c$ and even below when  a small magnetic field is applied, as shown in the log-log plot of Figure~\ref{rhoaPF6}-b. As pressure is increased, the linearity decreases in strength and the resistivity develops   some upward curvature which translates into      
a gradual increase of the power law exponent $\alpha$, but which still remains smaller than the Fermi liquid limit $\alpha=2$. It is only at much larger pressure ($P= 20.8$~kbar in Fig.~\ref{rhoaPF6}), when $T_c$ becomes small, that   the  $T^2$ behaviour tends to be retrieved as shown in Fig.~\ref{rhoaPF6}-b. Deviations to the Fermi liquid behaviour were found to scale with the size of $T_c$\cite{DoironLeyraud09,DoironLeyraud10}.

  These features are consistent with the results of Fig.~\ref{Alpha_T} where deviations from linearity develop with increasing the antinesting parameter so that the resistivity exponent reaches 2 when the superconducting phase is about to disappear. The calculated evolution of the exponent $\alpha$ near $t_\perp'^*$ and above  is in qualitative agreement with the results of Fig.~\ref{rhoaPF6}.

The theory, however, fails to predict the restoration  of  an apparent Fermi liquid $T^2$-behaviour in the intermediate temperature region, namely above the domain of anomalous resistivity.  As mentioned above, this Fermi liquid behaviour is also found  in the SDW sector of the phase diagram above $T_{\rm SDW} \sim 10$K (see Fig.~\ref{rhoaPF6}). In the present approach, this would indicate that the system  crossovers to a state where umklapp renormalization becomes  essentially absent,  in spite of reinforced nesting properties resulting from either a  temperature increase or a reduction of $t_\perp'$. These features are at variance  with  the results of the present model that invariably predicts the reverse. For systems like the Bechgaard salts with one type of carriers tied to a single component Fermi surface, these features may indicate that   a `colder' source of inelastic scattering\cite{Hlubina95}, not included in the present calculations,  gains in importance  at higher temperature. It  surpasses around 10K the anomalous power-law behaviour resulting from the sole basis of electronic half-filled  umklapp scattering. This is also inferred from   the calculated values of 3D resistivity    for  the present model. The actual  resistivity is obtained from the product of the 2D part by the interchain lattice constant $c$, namely $\rho_a c$. For the Bechgaard salts, $c=13.52$\AA  \cite{Jerome82} and the resistivity can then be expressed in $\mu \Omega\cdot\text{cm}$ units, as shown in the right-hand  side scale of Figures \ref{RhoaFT}-\ref{RhoSCd}. The comparison with the Fig.~\ref{rhoaPF6}-b for the inelastic part of resistivity indicates that  the data reach    one order of magnitude larger values at 10K, that is where the change of regime occurs (see Fig.~\ref{rhoaPF6}).

\vskip 1 cm 
\section{Summary  and concluding remarks}
We have studied the temperature and pressure dependence of electrical resistivity of  the Bechgaard salts  quasi-one dimensional organic superconductors in the vicinity of their quantum critical point where superconductivity borders on spin-density-wave order under pressure. We have proceeded to  the numerical solution of the semi-classical Boltzmann equation in which the collision umklapp electronic term has been computed by one-loop RG calculations   in the  framework of the quasi-1D electron gas model   in the presence of tunable nesting alterations. 

The temperature dependence of longitudinal resistivity $\rho_a$ has thus been obtained over the whole temperature domain from the 1D high temperature  region, which is invariably insulating, down to  the different effective 2D regimes where  either an  insulating SDW  or  SCd instabilities are  found depending on the strength of antinesting parameter $t_\perp'$. Antinesting proves to be the characteristic scale of the model for  weakening the growth of umklapp term under scaling, allowing  the onset of  anomalous  metallic   resistivity near the quantum critical point $t_\perp'^*$ along the antinesting axis. Marked deviations from the Fermi liquid  $T^2$-behaviour have  been found and described in terms of  power law resistivity $\rho_a \sim T^\alpha$ in the chain direction. Linear resistivity is found within a characteristic fan shape region of the phase diagram near $t_\perp'^*$, followed by a gradual increase of the  exponent $\alpha$, reaching the Fermi liquid limit at low enough temperature and sufficiently large antinesting  or  low superconducting $T_c$.

Quantum critical regime can be understood as  an anomalous increase of umklapp scattering with lowering temperature, an enhancement   that turns out to be  non uniform or anisotropic along the Fermi surface, with peaks at the positions where the Fermi velocity is purely longitudinal and nesting is  relatively poor. This umklapp vertex renormalization being directly linked to an enhancement of SDW  correlations, these can  be equivalently considered as the source of anomalous inelastic scattering. The region of non Fermi liquid values of $\alpha$ are found to extend  well beyond the confines of the QCP vicinity, a consequence of the reinforcement of umklapp scattering or SDW fluctuations by d-wave Cooper pairing. This leads to the apparent correlation of $\alpha$ with the size of $T_c$.  

The results have been shown to   qualitatively account for the linear-$T$ longitudinal resistivity observed near the QCP   of the Bechgaard salts phase diagram and for the gradual growth of $\alpha$ toward the Fermi liquid behaviour, as  pressure  moves away from the QCP  and $T_c$ is suppressed. 
However, discrepancies between theory and experiment remain.The present approach to resistivity fails to reproduce the crossover to an apparent Fermi liquid  in the intermediate temperature domain, namely above that of  quantum critical effects. For systems like the Bechgaard salts  with a simple open Fermi surface, this is likely to point to a mechanism of electron scattering that surperimposes to  that of half-filling umklapp with increasing  temperature.

 \acknowledgments
 The authors would like to thank H. Bakrim for his valuable comments on computing aspects of this work, and N. Doiron-Leyraud and P. Auban-Senzier for sharing their experimental data on the resistivity of the Bechgaard salts. 
 C. B. thanks the National Science and Engineering Research Council  of Canada (NSERC) and the R\'eseau Qu\'eb\'ecois des Mat\'eriaux de Pointe (RQMP) for financial support. Computational resources were provided
by the R\'eseau qu\'eb\'ecois de calcul de haute performance
(RQCHP) and Compute Canada.

\appendix

\section{The linearized Boltzmann  equation}
The linearized Boltzmann  equation is of the form 
\begin{equation}
 \sum\limits_{i,\bm{k^\prime}} \mathcal{L}^{[i]}_{\bm{k},\bm{k^\prime}}\phi_{\bm{k^\prime}}= e\beta \mathbf{\cal E}\cdot v_{\boldsymbol{k}},
 \label{LBEc}
\end{equation}
where the collision operator   given by expression (\ref{L1}) is the sum  of four terms 
\begin{widetext}
\begin{align}
\label{LBEa}
\sum\limits^4_{i=1} \mathcal{L}^{[i]}_{\bm{k},\bm{k^\prime}} = \dfrac{1}{(L N_P)^2} &\sum\limits_{\bm{k}_2,\bm{k}_3,\bm{k}_4}  {1\over 2} {\vert g_3(k_\perp,k_{\perp2},k_{\perp3},k_{\perp4}) - g_3(k_\perp,k_{\perp2},k_{\perp4},k_{\perp3}) \vert}^2
\times \frac{2\pi}{\hbar}  \delta_{\bm{k}+\bm{k}_2,\bm{k}_3+\bm{k}_4+ p \bm{G}}  \nonumber \\
&\times  \delta(\varepsilon^p_{\bm{k}}+\varepsilon^{p_2}_{\bm{k}_2}-\varepsilon^{p_3}_{\bm{k}_3}-\varepsilon^{p_4}_{\bm{k}_4}) \dfrac{f^0(\bm{k}_2)[1-f^0(\bm{k}_3)][1-f^0(\bm{k}_4)]}{[1-f^0(\bm{k})]} (\delta_{\bm{k},\bm{k^\prime}} + \delta_{\bm{k}_2,\bm{k^\prime}} - \delta_{\bm{k}_3,\bm{k^\prime}} - \delta_{\bm{k}_4,\bm{k^\prime}}).
\end{align}
\end{widetext} 
In  the framework of the electron gas model, we have used 
$$
\langle\bm{k}_1,\bm{k}_2\vert g_3 \vert\bm{k}_3,\bm{k}_4\rangle = \pi\hbar v_F g_3(k_{\perp1},k_{\perp2},k_{\perp3},k_{\perp4})
$$
for the (normalized) umklapp vertex functions  evaluated on the Fermi surface. By separating the kinematics constraint on the momentum conservation  into its longitudinal and transverse components\cite{Gorkov98}, one has 
\begin{align}
& \delta_{\bm{k}+\bm{k}_2,\bm{k}_3+\bm{k}_4+ p \bm{G}} =  \  \delta_{k_\perp + k_{\perp2},k_{\perp3}+k_{\perp4}} \cr
&  \ \ \ \ \times  \frac{2\pi}{L} \hbar v_F  \delta( \varepsilon^p_{\bm{k}}+\varepsilon^{p_2}_{\bm{k}_2}+\varepsilon^{p_3}_{\bm{k}_3}+\varepsilon^{p_4}_{\bm{k}_4} - \Sigma),\cr
\end{align}
where $\Sigma = \epsilon_\perp(k_\perp)+\epsilon_\perp(k_{\perp2})+\epsilon_\perp(k_{\perp3})+\epsilon_\perp(k_{\perp4})$.

The summation over the momentum vectors can be written as
\begin{eqnarray}
\label{sum}
\dfrac{1}{L N_P}\sum\limits_{\bm{k}} =  \sum_p \int \dfrac{d\varepsilon^p_{\bm{k}}}{2\pi \hbar v_F} \frac{1}{N_P} \sum\limits_{k_\perp}.
\end{eqnarray}
Carrying out the integration over the energy variables $(\varepsilon^{p_3}_{\bm{k}_3},\varepsilon^{p_4}_{\bm{k}_4}$) and  $(\varepsilon^{p_2}_{\bm{k}_2},\varepsilon^{p_4}_{\bm{k}_4}$) for respectively the first and the last two terms of (\ref{LBEa}), we arrive at the corresponding  expressions entering the collision operator,
\begin{widetext}
\begin{eqnarray}
\mathcal{L}_{\bm{k},\bm{k^\prime}}^{[1]} = \dfrac{\pi^2}{\hbar {N_P}^2} &\sum\limits_{k_{\perp3}, k_{\perp4}}& {\vert {g_3}(k_\perp , k_{\perp3}+k_{\perp4}-k_\perp , k_{\perp3}, k_{\perp4}) - {g_3}(k_\perp , k_{\perp3}+k_{\perp4}-k_\perp , k_{\perp4},k_{\perp 3}) \vert}^2 \nonumber \\
&\times & \dfrac{1+e^{-\beta \varepsilon^p_{\bm{k}}}}{1+e^{\beta(\Sigma / 2 - \varepsilon^p_{\bm{k}})}} \dfrac{\Sigma / 2 e^{\beta \Sigma / 2}}{e^{\beta \Sigma / 2}-1}  \delta_{\bm{k},\bm{k^\prime}}, \nonumber \\
\mathcal{L}^{[2]}_{\bm{k},\bm{k^\prime}} = \dfrac{\pi^3 v_F}{L {N_P}^2} &\sum\limits_{k_{\perp3}, k_{\perp4}}& {\vert {g_3}(k_\perp , k_{\perp}^\prime , k_{\perp3}, k_{\perp4}) - {g_3}(k_\perp , k^\prime_\perp , k_{\perp4},k_{\perp 3}) \vert}^2
 \nonumber \\
&\times &  \delta_{k_\perp + k^\prime_{\perp},k_{\perp3}+k_{\perp4}}  \delta(\varepsilon^p_{\bm{k}}+\varepsilon^{p'}_{\bm{k^\prime}} - \Sigma_1 /2) \dfrac{1+e^{-\beta \varepsilon^p_{\bm{k}}}}{1+e^{\beta \varepsilon^{p'}_{\bm{k^\prime}}}} \dfrac{(\varepsilon^p_{\bm{k}}+\varepsilon^{p'}_{\bm{k^\prime}}) e^{\beta (\varepsilon^{p}_{\bm{k}}+\varepsilon^{p'}_{\bm{k^\prime}})}}{e^{\beta (\varepsilon^p_{\bm{k}}+\varepsilon^{p'}_{\bm{k^\prime}})}-1},\nonumber \\
\mathcal{L}_{\bm{k},\bm{k^\prime}}^{[3]} = - \dfrac{\pi^3 v_F}{L {N_P}^2} &\sum\limits_{k_{\perp2}, k_{\perp4}}& {\vert {g_3}(k_\perp , k_{\perp2} , k_{\perp}^\prime,k_{\perp4}) - {g_3}(k_\perp , k_{\perp2} , k_{\perp4}, k_{\perp}^\prime) \vert}^2
  \nonumber \\
&\times &\delta_{k_\perp + k_{\perp2},k_{\perp}^\prime+k_{\perp4}} \dfrac{1+e^{-\beta \varepsilon^p_{\bm{k}}}}{1+e^{-\beta \varepsilon^{p'}_{\bm{k^\prime}}}} \dfrac{1}{(e^{\beta(\Sigma_2 /2-\varepsilon^p_{\bm{k}})}+1)(e^{-\beta(\Sigma_2 /2 -\varepsilon^{p'}_{\bm{k^\prime}})}+1)}, \nonumber \\
\mathcal{L}_{\bm{k},\bm{k^\prime}}^{[4]} = - \dfrac{\pi^3 v_F}{L {N_P}^2} &\sum\limits_{k_{\perp2}, k_{\perp3}}& {\vert {g_3}(k_\perp , k_{\perp2} , k_{\perp3},k^\prime_\perp) - {g_3}(k_\perp , k_{\perp2} , k^\prime_\perp,k_{\perp3}) \vert}^2
\nonumber \\
&\times & \delta_{k_\perp + k_{\perp2},k_{\perp3}+k_{\perp}^\prime}  \dfrac{1+e^{-\beta \varepsilon^p_{\bm{k}}}}{1+e^{-\beta \varepsilon^{p'}_{\bm{k^\prime}}}} \dfrac{1}{(e^{\beta(\Sigma_3 /2-\varepsilon^p_{\bm{k}})}+1)(e^{-\beta(\Sigma_3 /2 -\varepsilon^{p'}_{\bm{k^\prime}})}+1)}, 
\end{eqnarray}
\end{widetext}
where $\Sigma_1$, $\Sigma_2$, and $\Sigma_3$ are obtained by substituting respectively $k_{\perp2}$, $k_{\perp3}$, and $k_{\perp4}$ with $k_{\perp}^\prime$ in the $\Sigma$ expression.

The normalized equation to solve numerically for the $\phi's$ reads   
$$
\sum \limits_{i,\bm{k^\prime}}\bar{\mathcal{L}}_{\bm{k},\bm{k}^\prime}^{[i]}\bar{\phi}_{\bm{k}^\prime}=1,
$$
 where according to (\ref{LBEc}) we have defined the normalized quantities $\pi d_\perp \mathcal{L}^{[i]} / v_F\to \bar{\mathcal{L}}^{[i]}$ and $\phi/(e{\cal E}_a\beta d_\perp)\to \bar{\phi}$. Carrying out the integration over $\varepsilon^{p'}_{\bm{k^\prime}}$
and  rearranging the terms we arrive at the final equation,
\begin{widetext}
\begin{eqnarray}
\label{LBEb}
 {{\pi}^3d_\perp\over2a} { T\over E_F} {1\over N_P^2}\!  &\sum\limits_{k^\prime_{\perp},k_{\perp3}, k_{\perp4}}& \Big\{ {\vert {g_3}(k_\perp , k_{\perp3}+k_{\perp4}-k_\perp , k_{\perp3},k_{\perp4}) - {g_3}(k_\perp , k_{\perp3}+k_{\perp4}-k_\perp , k_{\perp4},k_{\perp3}) \vert}^2 \dfrac{\beta \Sigma' /2}{\sinh(\beta \Sigma' /2)} \delta_{k_\perp,k^\prime_\perp}  \nonumber \\
&+&  {\vert  {g_3}(k_\perp , k^\prime_\perp , k_{\perp3},k_{\perp4}) -  {g_3}(k_\perp , k^\prime_\perp , k_{\perp4}, k_{\perp3}) \vert}^2 \dfrac{\beta \Sigma^{\prime\prime} /2}{\sinh(\beta\Sigma^{\prime\prime} /2)} \delta_{k_\perp + k_{\perp}^\prime,k_{\perp3}+k_{\perp4}}  \nonumber \\
&-& 2 {\vert  {g_3}(k_\perp , k_{\perp3}, k_{\perp}^\prime, k_{\perp4}) - {g_3}(k_\perp , k_{\perp3} , k_{\perp4},k_{\perp}^\prime) \vert}^2 \dfrac{\beta \Sigma^{\prime\prime} /2}{\sinh(\beta \Sigma^{\prime\prime} /2)} \delta_{k_\perp + k_{\perp3}, k^\prime_\perp + k_{\perp4}} \Big\}  \bar{\phi}_{k^\prime_\perp} = 1,
\end{eqnarray}
\end{widetext}
where 
\begin{eqnarray}
\Sigma' &=& \epsilon_\perp(k_\perp) + \epsilon_\perp(k_{\perp3}+k_{\perp4}-k_\perp) + \epsilon_\perp(k_{\perp3}) + \epsilon_\perp(k_{\perp4}), \nonumber \\
\Sigma^{\prime\prime} &=& \epsilon_\perp(k_\perp) + \epsilon_\perp(k_\perp^\prime) + \epsilon_\perp(k_{\perp3}) + \epsilon_\perp(k_{\perp4}).
\end{eqnarray}
We shall fix the ratio $d_\perp/a \simeq 1.05$ in (\ref{LBEb})  from the known values of  lattice constants  in  the Bechgaard salts for the longitudinal  $\hat{a}$ (7.3\AA) and transverse $\hat{b}$ (7.7\AA) directions \cite{Jerome82}. The solution of (\ref{LBEb}) for the $\phi_{k_\perp}$ on the Fermi surface leads to the evaluation of conductivity (\ref{sigma}) as the inverse of resistivity.
\section{One-loop renormalization group for the electron gas }
\subsection{Vertices}
The renormalization group approach  to the quasi-1D electron gas is detailed in Refs\cite{Nickel06,Bourbon09,Sedeki12} .  In the outline, the method starts with the segmentation of  
 energy shells of thickness $\frac{1}{2}E_0(\ell)d\ell$,  located at $\pm \frac{1}{2}E_0(\ell)$ from either sides of the Fermi  sheets, into $N_p$ patches, each centred at a particular value of the transverse momentum $k_\perp$, where   $E_0(\ell)=E_0e^{-\ell}$ is the scaled bandwidth at  step $\ell$ and $E_0\equiv 2E_F$ is the initial bandwidth. The successive partial trace integration of electron degrees of freedom in the partition function leads to the renormalization or flow of the scattering amplitudes $g_i$ as a function of $\ell$. At the one-loop level, the flow combines  corrections from the electron-hole (Peierls) and electron-electron (Cooper) interfering channels of scattering 
and leads to the flow equations written in the compact form
\begin{widetext}
\begin{eqnarray}\label{Flowg}
&\partial_\ell g_{1}(k_{\perp1},k_{\perp 2},k_{\perp 3},k_{\perp 4})& = (-2 g_1 \circ g_1 + g_1 \circ g_2 + g_2 \circ g_1) \partial_\ell {\cal L}_P -(g_1 \circ g_2 + g_2 \circ g_1)\partial_\ell {\cal L}_C, \nonumber \\
&\partial_\ell g_{2}(k_{\perp1},k_{\perp 2},k_{\perp 3},k_{\perp 4})& = -(g_1 \circ g_1 +g_2 \circ g_2) \partial_\ell {\cal L}_C +(g_2 \circ g_2 + g_3 \circ g_3) \partial_\ell {\cal L}_P, \nonumber \\
&\partial_\ell g_{3}(k_{\perp1},k_{\perp 2},k_{\perp 3},k_{\perp 4})& = (- g_1 \circ g_3 - g_3 \circ g_1 + g_2 \circ g_3 + g_3 \circ g_2) \partial_\ell {\cal L}_P +2 g_2 \bullet g_3 \partial_\ell {\cal L}_P,
\end{eqnarray}
\end{widetext}
where $\partial_\ell= \partial/\partial \ell$. 
${\cal L}_{\nu=P,C}$ are the  Peierls and Cooper loops whose derivative at finite temperature comprises an integration over the patch. These   take the form
\begin{align}
\label{LPC}
 \partial_\ell {\cal L}_{\nu} & (k_\perp,q_{ \perp\nu}^{(\prime)})=  \frac{E_0(\ell)}{4N_P} \sum \limits_{\mu=\pm 1} \int_{k_\perp - \frac{\pi}{N_P}}^{k_\perp + \frac{\pi}{N_P}} {dk_\perp\over 2\pi} \cr
& \times \dfrac{\theta(\vert E_0(\ell)/2 + \mu A_\nu\vert - E_0(\ell)/2)}{E_0(\ell)+\mu A_\nu} \cr
&  \times   \Big[ \tanh[\beta E_0(\ell)/4] + \tanh[\beta( E_0(\ell)/4 + \mu A_\nu /2)]\Big],\cr 
\end{align}
where
\begin{align}
A_\nu(k_\perp,q_{\perp\nu}^{(\prime)})=&- \epsilon_\perp(k_\perp) - \eta_\nu \epsilon_\perp(\eta_\nu k_\perp+q_{\perp\nu}^{(\prime)})\cr
& + \eta_\nu \epsilon_\perp(\eta_\nu k_{\perp2(4)}+q_{ \perp\nu}^{(\prime)}) + \epsilon_\perp(k_{\perp2(4)})\cr
\end{align}
with $q_{\perp P}^{(\prime)}= k_{\perp3}-k_{\perp2}=k_{\perp1}-k_{\perp4} (k_{\perp3}-k_{\perp1}=k_{\perp2}-k_{\perp4})$ and $q_{\perp C}= k_{\perp1}+k_{\perp2}=k_{\perp3}+k_{\perp4} $; $\eta_P = 1$ and $\eta_C = -1$. $\theta(x)$ is the Heaviside function $[\theta(0) \equiv {1\over 2}]$.

The momentum dependence of couplings  in the discrete convolution  products  `$\circ$' over the internal $k_\perp$ loop variable on the right-hand side of (\ref{Flowg}) are in order $g(k_\perp,k_{\perp4},k_{\perp1}, k_\perp-q_{\perp P})g(k_\perp,k_{\perp2},k_{\perp3},k_\perp-q_{\perp P})$ for   the Peierls channel, $g(k_{\perp1},k_{\perp2},k_\perp,q_{\perp C}-k_\perp)g(k_{\perp3},k_{\perp4},k_\perp,q_{\perp C}-k_\perp)$ for the Cooper channel, and  $g(k_\perp,k_{\perp4},k_{\perp2},k_\perp-q_{\perp P}')g(k_{\perp1},k_\perp,k_{\perp3},k_{\perp1}-q_{\perp P}')$ for the `$\bullet$' product of the off-diagonal Peierls channel.

\subsection{Susceptibilities}
The  linear response of the electron system to a source field of SDW and SCD order parameter  leads to the corresponding expressions of the normalized static susceptibility
\begin{eqnarray}
\label{FlowKi}
{\bar{\chi}}_\mu(\bm{q}_\mu)=2 \int_\ell {(f_\mu(k_\perp)z_\mu(k_\perp))}^2 \partial_\ell \mathcal{L}_\mu d\ell
\end{eqnarray}
for $\mu=$SDW  at $\bm{q}_0=(2k_F,\pi)$ and $\mu=$SCd at $\bm{q}_0=0$.These are functions of  the  renormalization factor for the pair vertex $z_\mu$ in the corresponding channel which obeys the following flow equation
\begin{eqnarray}\label{FlowKiz}
\partial_\ell z_{\mu}(k_\perp)=(\partial_\ell \mathcal{L}_\mu)g_\mu z_\mu(k^\prime_\perp),
\end{eqnarray}
where
\begin{align}
\label{gmu}
 g_{\text{SDW}} = &\ g_2(k^\prime_\perp +\pi , k_\perp , k_\perp +\pi, k^\prime_\perp) \cr & +
 g_3(k^\prime_\perp -\pi , k_\perp +\pi , k_\perp,k^\prime_\perp),\cr  
 g_{\text{SCd}}  = & -g_1(-k^\prime_\perp , k^\prime_\perp , -k_\perp,k_\perp)\cr
 & -g_2(-k^\prime_\perp , k^\prime_\perp , -k_\perp,k_\perp), 
\end{align}
 are the combinations of momentum dependent scattering amplitudes that govern the flow for  each susceptibility. The form factors  $f_\mu$ assigned to the nature of ordered phases are $f_{\text{SDW}}=1$ and $f_{\text{SCd}}=\sqrt{2}\cos k_\perp $.

\bibliography{articlesII}

\begin{thebibliography}{47}%
\makeatletter
\providecommand \@ifxundefined [1]{%
 \@ifx{#1\undefined}
}%
\providecommand \@ifnum [1]{%
 \ifnum #1\expandafter \@firstoftwo
 \else \expandafter \@secondoftwo
 \fi
}%
\providecommand \@ifx [1]{%
 \ifx #1\expandafter \@firstoftwo
 \else \expandafter \@secondoftwo
 \fi
}%
\providecommand \natexlab [1]{#1}%
\providecommand \enquote  [1]{``#1''}%
\providecommand \bibnamefont  [1]{#1}%
\providecommand \bibfnamefont [1]{#1}%
\providecommand \citenamefont [1]{#1}%
\providecommand \href@noop [0]{\@secondoftwo}%
\providecommand \href [0]{\begingroup \@sanitize@url \@href}%
\providecommand \@href[1]{\@@startlink{#1}\@@href}%
\providecommand \@@href[1]{\endgroup#1\@@endlink}%
\providecommand \@sanitize@url [0]{\catcode `\\12\catcode `\$12\catcode
  `\&12\catcode `\#12\catcode `\^12\catcode `\_12\catcode `\%12\relax}%
\providecommand \@@startlink[1]{}%
\providecommand \@@endlink[0]{}%
\providecommand \url  [0]{\begingroup\@sanitize@url \@url }%
\providecommand \@url [1]{\endgroup\@href {#1}{\urlprefix }}%
\providecommand \urlprefix  [0]{URL }%
\providecommand \Eprint [0]{\href }%
\providecommand \doibase [0]{http://dx.doi.org/}%
\providecommand \selectlanguage [0]{\@gobble}%
\providecommand \bibinfo  [0]{\@secondoftwo}%
\providecommand \bibfield  [0]{\@secondoftwo}%
\providecommand \translation [1]{[#1]}%
\providecommand \BibitemOpen [0]{}%
\providecommand \bibitemStop [0]{}%
\providecommand \bibitemNoStop [0]{.\EOS\space}%
\providecommand \EOS [0]{\spacefactor3000\relax}%
\providecommand \BibitemShut  [1]{\csname bibitem#1\endcsname}%
\let\auto@bib@innerbib\@empty
\bibitem [{\citenamefont {J{\'e}rome}\ \emph {et~al.}(1980)\citenamefont
  {J{\'e}rome}, \citenamefont {Mazaud}, \citenamefont {Ribault},\ and\
  \citenamefont {Bechgaard}}]{Jerome80}%
  \BibitemOpen
  \bibfield  {author} {\bibinfo {author} {\bibfnamefont {D.}~\bibnamefont
  {J{\'e}rome}}, \bibinfo {author} {\bibfnamefont {A.}~\bibnamefont {Mazaud}},
  \bibinfo {author} {\bibfnamefont {M.}~\bibnamefont {Ribault}}, \ and\
  \bibinfo {author} {\bibfnamefont {K.}~\bibnamefont {Bechgaard}},\ }\href@noop
  {} {\bibfield  {journal} {\bibinfo  {journal} {J. Phys. (Paris) Lett.}\
  }\textbf {\bibinfo {volume} {41}},\ \bibinfo {pages} {L95} (\bibinfo {year}
  {1980})}\BibitemShut {NoStop}%
\bibitem [{\citenamefont {J{\'e}rome}\ and\ \citenamefont
  {Schulz}(1982)}]{Jerome82}%
  \BibitemOpen
  \bibfield  {author} {\bibinfo {author} {\bibfnamefont {D.}~\bibnamefont
  {J{\'e}rome}}\ and\ \bibinfo {author} {\bibfnamefont {H.~J.}\ \bibnamefont
  {Schulz}},\ }\href@noop {} {\bibfield  {journal} {\bibinfo  {journal} {Adv.
  Phys.}\ }\textbf {\bibinfo {volume} {31}},\ \bibinfo {pages} {299} (\bibinfo
  {year} {1982})}\BibitemShut {NoStop}%
\bibitem [{\citenamefont {Bourbonnais}\ and\ \citenamefont
  {J\'erome}(2008)}]{Bourbon08}%
  \BibitemOpen
  \bibfield  {author} {\bibinfo {author} {\bibfnamefont {C.}~\bibnamefont
  {Bourbonnais}}\ and\ \bibinfo {author} {\bibfnamefont {D.}~\bibnamefont
  {J\'erome}},\ }in\ \href@noop {} {\emph {\bibinfo {booktitle} {The Physics of
  Organic Superconductors and Conductors}}},\ Vol.\ \bibinfo {volume} {110,
  Springer Series in Materials Science},\ \bibinfo {editor} {edited by\
  \bibinfo {editor} {\bibfnamefont {A.}~\bibnamefont {Lebed}}}\ (\bibinfo
  {publisher} {Springer},\ \bibinfo {address} {Heidelberg},\ \bibinfo {year}
  {2008})\ p.\ \bibinfo {pages} {357},\ \bibinfo {note} {see also
  arXiv:cond-mat/0904.0617}\BibitemShut {NoStop}%
\bibitem [{\citenamefont {Brown}(2015)}]{Brown15}%
  \BibitemOpen
  \bibfield  {author} {\bibinfo {author} {\bibfnamefont {S.~E.}\ \bibnamefont
  {Brown}},\ }\href@noop {} {\bibfield  {journal} {\bibinfo  {journal} {Physica
  C}\ }\textbf {\bibinfo {volume} {514}},\ \bibinfo {pages} {279} (\bibinfo
  {year} {2015})}\BibitemShut {NoStop}%
\bibitem [{\citenamefont {Taillefer}(2010)}]{Taillefer10}%
  \BibitemOpen
  \bibfield  {author} {\bibinfo {author} {\bibfnamefont {L.}~\bibnamefont
  {Taillefer}},\ }\href@noop {} {\bibfield  {journal} {\bibinfo  {journal}
  {Annu. Rev. Condens. Matter Phys.}\ }\textbf {\bibinfo {volume} {1}},\
  \bibinfo {pages} {51} (\bibinfo {year} {2010})}\BibitemShut {NoStop}%
\bibitem [{\citenamefont {Armitage}\ \emph {et~al.}(2010)\citenamefont
  {Armitage}, \citenamefont {Fournier},\ and\ \citenamefont
  {Greene}}]{Armitage10}%
  \BibitemOpen
  \bibfield  {author} {\bibinfo {author} {\bibfnamefont {N.~P.}\ \bibnamefont
  {Armitage}}, \bibinfo {author} {\bibfnamefont {P.}~\bibnamefont {Fournier}},
  \ and\ \bibinfo {author} {\bibfnamefont {R.~L.}\ \bibnamefont {Greene}},\
  }\href@noop {} {\bibfield  {journal} {\bibinfo  {journal} {Rev. Mod. Phys.}\
  }\textbf {\bibinfo {volume} {82}},\ \bibinfo {pages} {2421} (\bibinfo {year}
  {2010})}\BibitemShut {NoStop}%
\bibitem [{\citenamefont {T.Tomita}\ \emph {et~al.}(2015)\citenamefont
  {T.Tomita}, \citenamefont {Kuga}, \citenamefont {Uwatoko}, \citenamefont
  {Coleman},\ and\ \citenamefont {Nakatsuji}}]{Tomita15}%
  \BibitemOpen
  \bibfield  {author} {\bibinfo {author} {\bibnamefont {T.Tomita}}, \bibinfo
  {author} {\bibfnamefont {K.}~\bibnamefont {Kuga}}, \bibinfo {author}
  {\bibfnamefont {Y.}~\bibnamefont {Uwatoko}}, \bibinfo {author} {\bibfnamefont
  {P.}~\bibnamefont {Coleman}}, \ and\ \bibinfo {author} {\bibfnamefont
  {S.}~\bibnamefont {Nakatsuji}},\ }\href@noop {} {\bibfield  {journal}
  {\bibinfo  {journal} {Science}\ }\textbf {\bibinfo {volume} {349}},\ \bibinfo
  {pages} {506} (\bibinfo {year} {2015})}\BibitemShut {NoStop}%
\bibitem [{\citenamefont {Scalapino}(2012)}]{Scalapino12}%
  \BibitemOpen
  \bibfield  {author} {\bibinfo {author} {\bibfnamefont {D.~J.}\ \bibnamefont
  {Scalapino}},\ }\href@noop {} {\bibfield  {journal} {\bibinfo  {journal}
  {Rev. Mod. Phys.}\ }\textbf {\bibinfo {volume} {84}},\ \bibinfo {pages}
  {1383} (\bibinfo {year} {2012})}\BibitemShut {NoStop}%
\bibitem [{\citenamefont {Stewart}(2011)}]{Stewart11}%
  \BibitemOpen
  \bibfield  {author} {\bibinfo {author} {\bibfnamefont {G.~R.}\ \bibnamefont
  {Stewart}},\ }\href@noop {} {\bibfield  {journal} {\bibinfo  {journal} {Rev.
  Mod. Phys.}\ }\textbf {\bibinfo {volume} {83}},\ \bibinfo {pages} {1589}
  (\bibinfo {year} {2011})}\BibitemShut {NoStop}%
\bibitem [{\citenamefont {Doiron-Leyraud}\ \emph {et~al.}(2009)\citenamefont
  {Doiron-Leyraud}, \citenamefont {Auban-Senzier}, \citenamefont {Ren\'e~de
  Cotret}, \citenamefont {Bourbonnais}, \citenamefont {J\'erome}, \citenamefont
  {Bechgaard},\ and\ \citenamefont {Taillefer}}]{DoironLeyraud09}%
  \BibitemOpen
  \bibfield  {author} {\bibinfo {author} {\bibfnamefont {N.}~\bibnamefont
  {Doiron-Leyraud}}, \bibinfo {author} {\bibfnamefont {P.}~\bibnamefont
  {Auban-Senzier}}, \bibinfo {author} {\bibfnamefont {S.}~\bibnamefont
  {Ren\'e~de Cotret}}, \bibinfo {author} {\bibfnamefont {C.}~\bibnamefont
  {Bourbonnais}}, \bibinfo {author} {\bibfnamefont {D.}~\bibnamefont
  {J\'erome}}, \bibinfo {author} {\bibfnamefont {K.}~\bibnamefont {Bechgaard}},
  \ and\ \bibinfo {author} {\bibfnamefont {L.}~\bibnamefont {Taillefer}},\
  }\href {\doibase 10.1103/PhysRevB.80.214531} {\bibfield  {journal} {\bibinfo
  {journal} {Phys. Rev. B}\ }\textbf {\bibinfo {volume} {80}},\ \bibinfo
  {pages} {214531} (\bibinfo {year} {2009})}\BibitemShut {NoStop}%
\bibitem [{\citenamefont {Doiron-Leyraud}\ \emph {et~al.}(2010)\citenamefont
  {Doiron-Leyraud}, \citenamefont {Auban-Senzier}, \citenamefont {de~Cotret},
  \citenamefont {Bourbonnais}, \citenamefont {Sedeki}, \citenamefont
  {J\'erome}, \citenamefont {Bechgaard},\ and\ \citenamefont
  {Taillefer}}]{DoironLeyraud10}%
  \BibitemOpen
  \bibfield  {author} {\bibinfo {author} {\bibfnamefont {N.}~\bibnamefont
  {Doiron-Leyraud}}, \bibinfo {author} {\bibfnamefont {P.}~\bibnamefont
  {Auban-Senzier}}, \bibinfo {author} {\bibfnamefont {S.~R.}\ \bibnamefont
  {de~Cotret}}, \bibinfo {author} {\bibfnamefont {C.}~\bibnamefont
  {Bourbonnais}}, \bibinfo {author} {\bibfnamefont {A.}~\bibnamefont {Sedeki}},
  \bibinfo {author} {\bibfnamefont {D.}~\bibnamefont {J\'erome}}, \bibinfo
  {author} {\bibfnamefont {K.}~\bibnamefont {Bechgaard}}, \ and\ \bibinfo
  {author} {\bibfnamefont {L.}~\bibnamefont {Taillefer}},\ }\href@noop {}
  {\bibfield  {journal} {\bibinfo  {journal} {Eur. Phys. J. B}\ }\textbf
  {\bibinfo {volume} {78}},\ \bibinfo {pages} {23} (\bibinfo {year}
  {2010})}\BibitemShut {NoStop}%
\bibitem [{\citenamefont {Gorkov}\ and\ \citenamefont
  {Mochena}(1998)}]{Gorkov98}%
  \BibitemOpen
  \bibfield  {author} {\bibinfo {author} {\bibfnamefont {L.~P.}\ \bibnamefont
  {Gorkov}}\ and\ \bibinfo {author} {\bibfnamefont {M.}~\bibnamefont
  {Mochena}},\ }\href@noop {} {\bibfield  {journal} {\bibinfo  {journal} {Phys.
  Rev. B}\ }\textbf {\bibinfo {volume} {57}},\ \bibinfo {pages} {6204}
  (\bibinfo {year} {1998})}\BibitemShut {NoStop}%
\bibitem [{\citenamefont {Zheleznyak}\ and\ \citenamefont
  {Yakovenko}(1999)}]{Zheleznyak99}%
  \BibitemOpen
  \bibfield  {author} {\bibinfo {author} {\bibfnamefont {A.~T.}\ \bibnamefont
  {Zheleznyak}}\ and\ \bibinfo {author} {\bibfnamefont {V.~M.}\ \bibnamefont
  {Yakovenko}},\ }\href@noop {} {\bibfield  {journal} {\bibinfo  {journal}
  {Eur. Phys. J. B}\ }\textbf {\bibinfo {volume} {11}},\ \bibinfo {pages} {385}
  (\bibinfo {year} {1999})}\BibitemShut {NoStop}%
\bibitem [{\citenamefont {Emery}\ \emph {et~al.}(1982)\citenamefont {Emery},
  \citenamefont {Bruinsma},\ and\ \citenamefont {Barisic}}]{Emery82}%
  \BibitemOpen
  \bibfield  {author} {\bibinfo {author} {\bibfnamefont {V.~J.}\ \bibnamefont
  {Emery}}, \bibinfo {author} {\bibfnamefont {R.}~\bibnamefont {Bruinsma}}, \
  and\ \bibinfo {author} {\bibfnamefont {S.}~\bibnamefont {Barisic}},\
  }\href@noop {} {\bibfield  {journal} {\bibinfo  {journal} {Phys. Rev. Lett.}\
  }\textbf {\bibinfo {volume} {48}},\ \bibinfo {pages} {1039} (\bibinfo {year}
  {1982})}\BibitemShut {NoStop}%
\bibitem [{\citenamefont {Bourbonnais}\ \emph {et~al.}(1984)\citenamefont
  {Bourbonnais}, \citenamefont {Creuzet}, \citenamefont {J{\'e}rome},
  \citenamefont {Bechgaard},\ and\ \citenamefont {Moradpour}}]{Bourbon84}%
  \BibitemOpen
  \bibfield  {author} {\bibinfo {author} {\bibfnamefont {C.}~\bibnamefont
  {Bourbonnais}}, \bibinfo {author} {\bibfnamefont {F.}~\bibnamefont
  {Creuzet}}, \bibinfo {author} {\bibfnamefont {D.}~\bibnamefont {J{\'e}rome}},
  \bibinfo {author} {\bibfnamefont {K.}~\bibnamefont {Bechgaard}}, \ and\
  \bibinfo {author} {\bibfnamefont {A.}~\bibnamefont {Moradpour}},\ }\href@noop
  {} {\bibfield  {journal} {\bibinfo  {journal} {J. Phys. (Paris) Lett.}\
  }\textbf {\bibinfo {volume} {45}},\ \bibinfo {pages} {L755} (\bibinfo {year}
  {1984})}\BibitemShut {NoStop}%
\bibitem [{\citenamefont {Creuzet}\ \emph {et~al.}(1987)\citenamefont
  {Creuzet}, \citenamefont {Bourbonnais}, \citenamefont {Caron}, \citenamefont
  {J{\'e}rome},\ and\ \citenamefont {Moradpour}}]{Creuzet87b}%
  \BibitemOpen
  \bibfield  {author} {\bibinfo {author} {\bibfnamefont {F.}~\bibnamefont
  {Creuzet}}, \bibinfo {author} {\bibfnamefont {C.}~\bibnamefont
  {Bourbonnais}}, \bibinfo {author} {\bibfnamefont {L.~G.}\ \bibnamefont
  {Caron}}, \bibinfo {author} {\bibfnamefont {D.}~\bibnamefont {J{\'e}rome}}, \
  and\ \bibinfo {author} {\bibfnamefont {A.}~\bibnamefont {Moradpour}},\
  }\href@noop {} {\bibfield  {journal} {\bibinfo  {journal} {Synth. Met.}\
  }\textbf {\bibinfo {volume} {19}},\ \bibinfo {pages} {277} (\bibinfo {year}
  {1987})}\BibitemShut {NoStop}%
\bibitem [{\citenamefont {Wu}\ \emph {et~al.}(2005)\citenamefont {Wu},
  \citenamefont {Chaikin}, \citenamefont {Kang}, \citenamefont {Shinagawa},
  \citenamefont {Yu},\ and\ \citenamefont {Brown}}]{Wu05}%
  \BibitemOpen
  \bibfield  {author} {\bibinfo {author} {\bibfnamefont {W.}~\bibnamefont
  {Wu}}, \bibinfo {author} {\bibfnamefont {P.~M.}\ \bibnamefont {Chaikin}},
  \bibinfo {author} {\bibfnamefont {W.}~\bibnamefont {Kang}}, \bibinfo {author}
  {\bibfnamefont {J.}~\bibnamefont {Shinagawa}}, \bibinfo {author}
  {\bibfnamefont {W.}~\bibnamefont {Yu}}, \ and\ \bibinfo {author}
  {\bibfnamefont {S.~E.}\ \bibnamefont {Brown}},\ }\href@noop {} {\bibfield
  {journal} {\bibinfo  {journal} {Phys. Rev. Lett.}\ }\textbf {\bibinfo
  {volume} {94}},\ \bibinfo {pages} {097004} (\bibinfo {year}
  {2005})}\BibitemShut {NoStop}%
\bibitem [{\citenamefont {Brown}\ \emph {et~al.}(2008)\citenamefont {Brown},
  \citenamefont {Chaikin},\ and\ \citenamefont {Naughton}}]{Brown08}%
  \BibitemOpen
  \bibfield  {author} {\bibinfo {author} {\bibfnamefont {S.~E.}\ \bibnamefont
  {Brown}}, \bibinfo {author} {\bibfnamefont {P.~M.}\ \bibnamefont {Chaikin}},
  \ and\ \bibinfo {author} {\bibfnamefont {M.~J.}\ \bibnamefont {Naughton}},\
  }in\ \href@noop {} {\emph {\bibinfo {booktitle} {The Physics of Organic
  Superconductors and Conductors}}},\ Vol.\ \bibinfo {volume} {110, Springer
  Series in Materials Science},\ \bibinfo {editor} {edited by\ \bibinfo
  {editor} {\bibfnamefont {A.}~\bibnamefont {Lebed}}}\ (\bibinfo  {publisher}
  {Springer},\ \bibinfo {address} {Heidelberg},\ \bibinfo {year} {2008})\
  p.~\bibinfo {pages} {49}\BibitemShut {NoStop}%
\bibitem [{\citenamefont {Kimura}\ \emph {et~al.}(2011)\citenamefont {Kimura},
  \citenamefont {Misawa},\ and\ \citenamefont {Kawamoto}}]{Kimura11}%
  \BibitemOpen
  \bibfield  {author} {\bibinfo {author} {\bibfnamefont {Y.}~\bibnamefont
  {Kimura}}, \bibinfo {author} {\bibfnamefont {M.}~\bibnamefont {Misawa}}, \
  and\ \bibinfo {author} {\bibfnamefont {A.}~\bibnamefont {Kawamoto}},\
  }\href@noop {} {\bibfield  {journal} {\bibinfo  {journal} {Phys. Rev. B}\
  }\textbf {\bibinfo {volume} {84}},\ \bibinfo {pages} {045123} (\bibinfo
  {year} {2011})}\BibitemShut {NoStop}%
\bibitem [{\citenamefont {Bourbonnais}\ and\ \citenamefont
  {Sedeki}(2009)}]{Bourbon09}%
  \BibitemOpen
  \bibfield  {author} {\bibinfo {author} {\bibfnamefont {C.}~\bibnamefont
  {Bourbonnais}}\ and\ \bibinfo {author} {\bibfnamefont {A.}~\bibnamefont
  {Sedeki}},\ }\href@noop {} {\bibfield  {journal} {\bibinfo  {journal} {Phys.
  Rev. B}\ }\textbf {\bibinfo {volume} {80}},\ \bibinfo {pages} {085105}
  (\bibinfo {year} {2009})}\BibitemShut {NoStop}%
\bibitem [{\citenamefont {Nickel}\ \emph {et~al.}(2006)\citenamefont {Nickel},
  \citenamefont {Duprat}, \citenamefont {Bourbonnais},\ and\ \citenamefont
  {Dupuis}}]{Nickel06}%
  \BibitemOpen
  \bibfield  {author} {\bibinfo {author} {\bibfnamefont {J.~C.}\ \bibnamefont
  {Nickel}}, \bibinfo {author} {\bibfnamefont {R.}~\bibnamefont {Duprat}},
  \bibinfo {author} {\bibfnamefont {C.}~\bibnamefont {Bourbonnais}}, \ and\
  \bibinfo {author} {\bibfnamefont {N.}~\bibnamefont {Dupuis}},\ }\href@noop {}
  {\bibfield  {journal} {\bibinfo  {journal} {Phys. Rev. B}\ }\textbf {\bibinfo
  {volume} {73}},\ \bibinfo {pages} {165126} (\bibinfo {year}
  {2006})}\BibitemShut {NoStop}%
\bibitem [{\citenamefont {Sedeki}\ \emph {et~al.}(2012)\citenamefont {Sedeki},
  \citenamefont {Bergeron},\ and\ \citenamefont {Bourbonnais}}]{Sedeki12}%
  \BibitemOpen
  \bibfield  {author} {\bibinfo {author} {\bibfnamefont {A.}~\bibnamefont
  {Sedeki}}, \bibinfo {author} {\bibfnamefont {D.}~\bibnamefont {Bergeron}}, \
  and\ \bibinfo {author} {\bibfnamefont {C.}~\bibnamefont {Bourbonnais}},\
  }\href@noop {} {\bibfield  {journal} {\bibinfo  {journal} {Phys. Rev. B}\
  }\textbf {\bibinfo {volume} {85}},\ \bibinfo {pages} {165129} (\bibinfo
  {year} {2012})}\BibitemShut {NoStop}%
\bibitem [{\citenamefont {Meier}\ \emph {et~al.}(2013)\citenamefont {Meier},
  \citenamefont {Auban-Senzier}, \citenamefont {P\'epin},\ and\ \citenamefont
  {J\'erome}}]{Meier13}%
  \BibitemOpen
  \bibfield  {author} {\bibinfo {author} {\bibfnamefont {H.}~\bibnamefont
  {Meier}}, \bibinfo {author} {\bibfnamefont {P.}~\bibnamefont
  {Auban-Senzier}}, \bibinfo {author} {\bibfnamefont {C.}~\bibnamefont
  {P\'epin}}, \ and\ \bibinfo {author} {\bibfnamefont {D.}~\bibnamefont
  {J\'erome}},\ }\href@noop {} {\bibfield  {journal} {\bibinfo  {journal}
  {Phys. Rev. B}\ }\textbf {\bibinfo {volume} {87}},\ \bibinfo {pages} {125128}
  (\bibinfo {year} {2013})}\BibitemShut {NoStop}%
\bibitem [{\citenamefont {Buhmann}\ \emph {et~al.}(2013)\citenamefont
  {Buhmann}, \citenamefont {Ossadnik}, \citenamefont {Rice},\ and\
  \citenamefont {Sigrist}}]{Buhmann13}%
  \BibitemOpen
  \bibfield  {author} {\bibinfo {author} {\bibfnamefont {J.~M.}\ \bibnamefont
  {Buhmann}}, \bibinfo {author} {\bibfnamefont {M.}~\bibnamefont {Ossadnik}},
  \bibinfo {author} {\bibfnamefont {T.~M.}\ \bibnamefont {Rice}}, \ and\
  \bibinfo {author} {\bibfnamefont {M.}~\bibnamefont {Sigrist}},\ }\href@noop
  {} {\bibfield  {journal} {\bibinfo  {journal} {Phys. Rev. B}\ }\textbf
  {\bibinfo {volume} {87}},\ \bibinfo {pages} {035129} (\bibinfo {year}
  {2013})}\BibitemShut {NoStop}%
\bibitem [{\citenamefont {Haug}\ and\ \citenamefont {Jauho}(2008)}]{Haug08}%
  \BibitemOpen
  \bibfield  {author} {\bibinfo {author} {\bibfnamefont {H.}~\bibnamefont
  {Haug}}\ and\ \bibinfo {author} {\bibfnamefont {A.~P.}\ \bibnamefont
  {Jauho}},\ }in\ \href@noop {} {\emph {\bibinfo {booktitle} {Quantum Kinetics
  in Transport and Optics of Semiconductors}}},\ Vol.\ \bibinfo {volume} {110,
  Springer Series in Solid-State Sciences},\ \bibinfo {editor} {edited by\
  \bibinfo {editor} {\bibfnamefont {M.}~\bibnamefont {Cardona}}\ and\ \bibinfo
  {editor} {\bibfnamefont {P.}~\bibnamefont {Fulde}}}\ (\bibinfo  {publisher}
  {Springer},\ \bibinfo {address} {Heidelberg},\ \bibinfo {year} {2008})\
  p.~\bibinfo {pages} {3}\BibitemShut {NoStop}%
\bibitem [{\citenamefont {Wzietek}\ \emph {et~al.}(1993)\citenamefont
  {Wzietek}, \citenamefont {Creuzet}, \citenamefont {Bourbonnais},
  \citenamefont {J{\'e}rome}, \citenamefont {Bechgaard},\ and\ \citenamefont
  {Batail}}]{Wzietek93}%
  \BibitemOpen
  \bibfield  {author} {\bibinfo {author} {\bibfnamefont {P.}~\bibnamefont
  {Wzietek}}, \bibinfo {author} {\bibfnamefont {F.}~\bibnamefont {Creuzet}},
  \bibinfo {author} {\bibfnamefont {C.}~\bibnamefont {Bourbonnais}}, \bibinfo
  {author} {\bibfnamefont {D.}~\bibnamefont {J{\'e}rome}}, \bibinfo {author}
  {\bibfnamefont {K.}~\bibnamefont {Bechgaard}}, \ and\ \bibinfo {author}
  {\bibfnamefont {P.}~\bibnamefont {Batail}},\ }\href@noop {} {\bibfield
  {journal} {\bibinfo  {journal} {J. Phys. I (France)}\ }\textbf {\bibinfo
  {volume} {3}},\ \bibinfo {pages} {171} (\bibinfo {year} {1993})}\BibitemShut
  {NoStop}%
\bibitem [{\citenamefont {Klemme}\ \emph {et~al.}(1995)\citenamefont {Klemme},
  \citenamefont {Brown}, \citenamefont {Wzietek}, \citenamefont {G.~Kriza},
  \citenamefont {J{\'e}rome},\ and\ \citenamefont {Fabre}}]{Klemme95}%
  \BibitemOpen
  \bibfield  {author} {\bibinfo {author} {\bibfnamefont {B.~J.}\ \bibnamefont
  {Klemme}}, \bibinfo {author} {\bibfnamefont {S.~E.}\ \bibnamefont {Brown}},
  \bibinfo {author} {\bibfnamefont {P.}~\bibnamefont {Wzietek}}, \bibinfo
  {author} {\bibfnamefont {P.~B.}\ \bibnamefont {G.~Kriza}}, \bibinfo {author}
  {\bibfnamefont {D.}~\bibnamefont {J{\'e}rome}}, \ and\ \bibinfo {author}
  {\bibfnamefont {J.-M.}\ \bibnamefont {Fabre}},\ }\href@noop {} {\bibfield
  {journal} {\bibinfo  {journal} {Phys. Rev. Lett.}\ }\textbf {\bibinfo
  {volume} {75}},\ \bibinfo {pages} {2408} (\bibinfo {year}
  {1995})}\BibitemShut {NoStop}%
\bibitem [{\citenamefont {Moser}\ \emph {et~al.}(1998)\citenamefont {Moser},
  \citenamefont {Gabay}, \citenamefont {Auban-Senzier}, \citenamefont
  {J{\'e}rome}, \citenamefont {Bechgaard},\ and\ \citenamefont
  {Fabre}}]{Moser98}%
  \BibitemOpen
  \bibfield  {author} {\bibinfo {author} {\bibfnamefont {J.}~\bibnamefont
  {Moser}}, \bibinfo {author} {\bibfnamefont {M.}~\bibnamefont {Gabay}},
  \bibinfo {author} {\bibfnamefont {P.}~\bibnamefont {Auban-Senzier}}, \bibinfo
  {author} {\bibfnamefont {D.}~\bibnamefont {J{\'e}rome}}, \bibinfo {author}
  {\bibfnamefont {K.}~\bibnamefont {Bechgaard}}, \ and\ \bibinfo {author}
  {\bibfnamefont {J.~M.}\ \bibnamefont {Fabre}},\ }\href@noop {} {\bibfield
  {journal} {\bibinfo  {journal} {Eur. Phys. J. B}\ }\textbf {\bibinfo {volume}
  {1}},\ \bibinfo {pages} {39} (\bibinfo {year} {1998})}\BibitemShut {NoStop}%
\bibitem [{\citenamefont {Grant}(1983)}]{Grant83}%
  \BibitemOpen
  \bibfield  {author} {\bibinfo {author} {\bibfnamefont {P.~M.}\ \bibnamefont
  {Grant}},\ }\href@noop {} {\bibfield  {journal} {\bibinfo  {journal} {J.
  Phys. (Paris) Coll.}\ }\textbf {\bibinfo {volume} {44}},\ \bibinfo {pages}
  {847} (\bibinfo {year} {1983})}\BibitemShut {NoStop}%
\bibitem [{\citenamefont {Barisic}\ and\ \citenamefont
  {Brazovskii}(1981)}]{Barisic81}%
  \BibitemOpen
  \bibfield  {author} {\bibinfo {author} {\bibfnamefont {S.}~\bibnamefont
  {Barisic}}\ and\ \bibinfo {author} {\bibfnamefont {S.}~\bibnamefont
  {Brazovskii}},\ }in\ \href@noop {} {\emph {\bibinfo {booktitle} {Recent
  Developments in Condensed Matter Physics}}},\ Vol.~\bibinfo {volume} {1},\
  \bibinfo {editor} {edited by\ \bibinfo {editor} {\bibfnamefont {J.~T.}\
  \bibnamefont {Devreese}}}\ (\bibinfo  {publisher} {Plenum},\ \bibinfo
  {address} {New York},\ \bibinfo {year} {1981})\ p.\ \bibinfo {pages}
  {327}\BibitemShut {NoStop}%
\bibitem [{\citenamefont {Mila}\ and\ \citenamefont {Penc}(1994)}]{Mila94}%
  \BibitemOpen
  \bibfield  {author} {\bibinfo {author} {\bibfnamefont {F.}~\bibnamefont
  {Mila}}\ and\ \bibinfo {author} {\bibfnamefont {K.}~\bibnamefont {Penc}},\
  }\href@noop {} {\bibfield  {journal} {\bibinfo  {journal} {Phys. Rev. B}\
  }\textbf {\bibinfo {volume} {50}},\ \bibinfo {pages} {11429} (\bibinfo {year}
  {1994})}\BibitemShut {NoStop}%
\bibitem [{\citenamefont {Grant}(1982)}]{Grant82}%
  \BibitemOpen
  \bibfield  {author} {\bibinfo {author} {\bibfnamefont {P.~M.}\ \bibnamefont
  {Grant}},\ }\href@noop {} {\bibfield  {journal} {\bibinfo  {journal} {Phys.
  Rev. B}\ }\textbf {\bibinfo {volume} {26}},\ \bibinfo {pages} {6888}
  (\bibinfo {year} {1982})}\BibitemShut {NoStop}%
\bibitem [{\citenamefont {Ducasse}\ \emph {et~al.}(1985)\citenamefont
  {Ducasse}, \citenamefont {Abderrabba},\ and\ \citenamefont
  {Gallois}}]{Ducasse85}%
  \BibitemOpen
  \bibfield  {author} {\bibinfo {author} {\bibfnamefont {L.}~\bibnamefont
  {Ducasse}}, \bibinfo {author} {\bibfnamefont {A.}~\bibnamefont {Abderrabba}},
  \ and\ \bibinfo {author} {\bibfnamefont {B.}~\bibnamefont {Gallois}},\
  }\href@noop {} {\bibfield  {journal} {\bibinfo  {journal} {J. Phys. C}\
  }\textbf {\bibinfo {volume} {18}},\ \bibinfo {pages} {L947} (\bibinfo {year}
  {1985})}\BibitemShut {NoStop}%
\bibitem [{\citenamefont {Bourbonnais}\ and\ \citenamefont
  {Caron}(1991)}]{Bourbon91}%
  \BibitemOpen
  \bibfield  {author} {\bibinfo {author} {\bibfnamefont {C.}~\bibnamefont
  {Bourbonnais}}\ and\ \bibinfo {author} {\bibfnamefont {L.~G.}\ \bibnamefont
  {Caron}},\ }\href@noop {} {\bibfield  {journal} {\bibinfo  {journal} {Int. J.
  Mod. Phys. B}\ }\textbf {\bibinfo {volume} {05}},\ \bibinfo {pages} {1033}
  (\bibinfo {year} {1991})}\BibitemShut {NoStop}%
\bibitem [{\citenamefont {Bourbonnais}\ \emph {et~al.}(2003)\citenamefont
  {Bourbonnais}, \citenamefont {Guay},\ and\ \citenamefont
  {Wortis}}]{Bourbon03}%
  \BibitemOpen
  \bibfield  {author} {\bibinfo {author} {\bibfnamefont {C.}~\bibnamefont
  {Bourbonnais}}, \bibinfo {author} {\bibfnamefont {B.}~\bibnamefont {Guay}}, \
  and\ \bibinfo {author} {\bibfnamefont {R.}~\bibnamefont {Wortis}},\ }in\
  \href@noop {} {\emph {\bibinfo {booktitle} {Theoretical methods for strongly
  correlated electrons}}},\ \bibinfo {editor} {edited by\ \bibinfo {editor}
  {\bibfnamefont {D.}~\bibnamefont {S\'en\'echal}}, \bibinfo {editor}
  {\bibfnamefont {A.~M.}\ \bibnamefont {Tremblay}}, \ and\ \bibinfo {editor}
  {\bibfnamefont {C.}~\bibnamefont {Bourbonnais}}}\ (\bibinfo  {publisher}
  {Springer},\ \bibinfo {address} {Heidelberg},\ \bibinfo {year} {2003})\ pp.\
  \bibinfo {pages} {77--78},\ \bibinfo {note} {arXiv:
  cond-mat/0204163}\BibitemShut {NoStop}%
\bibitem [{\citenamefont {Bourbonnais}\ and\ \citenamefont
  {Sedeki}(2011)}]{Bourbon11}%
  \BibitemOpen
  \bibfield  {author} {\bibinfo {author} {\bibfnamefont {C.}~\bibnamefont
  {Bourbonnais}}\ and\ \bibinfo {author} {\bibfnamefont {A.}~\bibnamefont
  {Sedeki}},\ }\href@noop {} {\bibfield  {journal} {\bibinfo  {journal} {C. R.
  Physique}\ }\textbf {\bibinfo {volume} {12}},\ \bibinfo {pages} {532}
  (\bibinfo {year} {2011})}\BibitemShut {NoStop}%
\bibitem [{\citenamefont {Gor'kov}\ and\ \citenamefont
  {Dzyaloshinskii}(1973)}]{Gorkov73}%
  \BibitemOpen
  \bibfield  {author} {\bibinfo {author} {\bibfnamefont {L.~P.}\ \bibnamefont
  {Gor'kov}}\ and\ \bibinfo {author} {\bibfnamefont {I.~E.}\ \bibnamefont
  {Dzyaloshinskii}},\ }\href@noop {} {\bibfield  {journal} {\bibinfo  {journal}
  {JETP Lett.}\ }\textbf {\bibinfo {volume} {18}},\ \bibinfo {pages} {401}
  (\bibinfo {year} {1973})}\BibitemShut {NoStop}%
\bibitem [{\citenamefont {Giamarchi}(1991)}]{Giamarchi91}%
  \BibitemOpen
  \bibfield  {author} {\bibinfo {author} {\bibfnamefont {T.}~\bibnamefont
  {Giamarchi}},\ }\href@noop {} {\bibfield  {journal} {\bibinfo  {journal}
  {Phys. Rev. B}\ }\textbf {\bibinfo {volume} {44}},\ \bibinfo {pages} {2905}
  (\bibinfo {year} {1991})}\BibitemShut {NoStop}%
\bibitem [{\citenamefont {Moriya}\ and\ \citenamefont {Ueda}(2003)}]{Moriya03}%
  \BibitemOpen
  \bibfield  {author} {\bibinfo {author} {\bibfnamefont {T.}~\bibnamefont
  {Moriya}}\ and\ \bibinfo {author} {\bibfnamefont {K.}~\bibnamefont {Ueda}},\
  }\href@noop {} {\bibfield  {journal} {\bibinfo  {journal} {Rep. Prog. Phys.}\
  }\textbf {\bibinfo {volume} {66}},\ \bibinfo {pages} {1299} (\bibinfo {year}
  {2003})}\BibitemShut {NoStop}%
\bibitem [{\citenamefont {Bechgaard}\ \emph {et~al.}(1980)\citenamefont
  {Bechgaard}, \citenamefont {Jacobsen}, \citenamefont {Mortensen},
  \citenamefont {Pedersen},\ and\ \citenamefont {Thorup}}]{Bechgaard80}%
  \BibitemOpen
  \bibfield  {author} {\bibinfo {author} {\bibfnamefont {K.}~\bibnamefont
  {Bechgaard}}, \bibinfo {author} {\bibfnamefont {C.}~\bibnamefont {Jacobsen}},
  \bibinfo {author} {\bibfnamefont {K.}~\bibnamefont {Mortensen}}, \bibinfo
  {author} {\bibfnamefont {H.}~\bibnamefont {Pedersen}}, \ and\ \bibinfo
  {author} {\bibfnamefont {N.}~\bibnamefont {Thorup}},\ }\href@noop {}
  {\bibfield  {journal} {\bibinfo  {journal} {Solid State Comm.}\ }\textbf
  {\bibinfo {volume} {33}},\ \bibinfo {pages} {1119} (\bibinfo {year}
  {1980})}\BibitemShut {NoStop}%
\bibitem [{\citenamefont {Vuletic}\ \emph {et~al.}(2002)\citenamefont
  {Vuletic}, \citenamefont {Auban-Senzier}, \citenamefont {Pasquier},
  \citenamefont {Tomic}, \citenamefont {Jerome}, \citenamefont {Heritier},\
  and\ \citenamefont {Bechgaard}}]{Vuletic02}%
  \BibitemOpen
  \bibfield  {author} {\bibinfo {author} {\bibfnamefont {T.}~\bibnamefont
  {Vuletic}}, \bibinfo {author} {\bibfnamefont {P.}~\bibnamefont
  {Auban-Senzier}}, \bibinfo {author} {\bibfnamefont {C.}~\bibnamefont
  {Pasquier}}, \bibinfo {author} {\bibfnamefont {S.}~\bibnamefont {Tomic}},
  \bibinfo {author} {\bibfnamefont {D.}~\bibnamefont {Jerome}}, \bibinfo
  {author} {\bibfnamefont {M.}~\bibnamefont {Heritier}}, \ and\ \bibinfo
  {author} {\bibfnamefont {K.}~\bibnamefont {Bechgaard}},\ }\href@noop {}
  {\bibfield  {journal} {\bibinfo  {journal} {Eur. Phys. J. B}\ }\textbf
  {\bibinfo {volume} {25}},\ \bibinfo {pages} {319} (\bibinfo {year}
  {2002})}\BibitemShut {NoStop}%
\bibitem [{\citenamefont {Coulon}\ \emph {et~al.}(1982)\citenamefont {Coulon},
  \citenamefont {Delhaes}, \citenamefont {Flandrois}, \citenamefont {Lagnier},
  \citenamefont {Bonjour},\ and\ \citenamefont {Fabre}}]{Coulon82}%
  \BibitemOpen
  \bibfield  {author} {\bibinfo {author} {\bibfnamefont {C.}~\bibnamefont
  {Coulon}}, \bibinfo {author} {\bibfnamefont {P.}~\bibnamefont {Delhaes}},
  \bibinfo {author} {\bibfnamefont {S.}~\bibnamefont {Flandrois}}, \bibinfo
  {author} {\bibfnamefont {R.}~\bibnamefont {Lagnier}}, \bibinfo {author}
  {\bibfnamefont {E.}~\bibnamefont {Bonjour}}, \ and\ \bibinfo {author}
  {\bibfnamefont {J.}~\bibnamefont {Fabre}},\ }\href@noop {} {\bibfield
  {journal} {\bibinfo  {journal} {J. Phys. (Paris)}\ }\textbf {\bibinfo
  {volume} {43}},\ \bibinfo {pages} {1059} (\bibinfo {year}
  {1982})}\BibitemShut {NoStop}%
\bibitem [{\citenamefont {Balicas}\ \emph {et~al.}(1994)\citenamefont
  {Balicas}, \citenamefont {Benhia}, \citenamefont {Kang}, \citenamefont
  {Canadell}, \citenamefont {P.~Auban-Senzier},\ and\ \citenamefont
  {Fabre}}]{Balicas94}%
  \BibitemOpen
  \bibfield  {author} {\bibinfo {author} {\bibfnamefont {L.}~\bibnamefont
  {Balicas}}, \bibinfo {author} {\bibfnamefont {K.}~\bibnamefont {Benhia}},
  \bibinfo {author} {\bibfnamefont {W.}~\bibnamefont {Kang}}, \bibinfo {author}
  {\bibfnamefont {E.}~\bibnamefont {Canadell}}, \bibinfo {author}
  {\bibfnamefont {M.~R.}\ \bibnamefont {P.~Auban-Senzier}, \bibfnamefont
  {D.~J{\'e}rome}}, \ and\ \bibinfo {author} {\bibfnamefont {J.}~\bibnamefont
  {Fabre}},\ }\href@noop {} {\bibfield  {journal} {\bibinfo  {journal} {J.
  Phys. I (France)}\ }\textbf {\bibinfo {volume} {4}},\ \bibinfo {pages} {1539}
  (\bibinfo {year} {1994})}\BibitemShut {NoStop}%
\bibitem [{\citenamefont {Giamarchi}(1997)}]{Giamarchi97}%
  \BibitemOpen
  \bibfield  {author} {\bibinfo {author} {\bibfnamefont {T.}~\bibnamefont
  {Giamarchi}},\ }\href@noop {} {\bibfield  {journal} {\bibinfo  {journal}
  {Physica}\ }\textbf {\bibinfo {volume} {B230-232}},\ \bibinfo {pages} {975}
  (\bibinfo {year} {1997})}\BibitemShut {NoStop}%
\bibitem [{\citenamefont {Schwartz}\ \emph {et~al.}(1998)\citenamefont
  {Schwartz}, \citenamefont {Dressel}, \citenamefont {Gr\"uner}, \citenamefont
  {Vescoli}, \citenamefont {Degiorgi},\ and\ \citenamefont
  {Giamarchi}}]{Schwartz98}%
  \BibitemOpen
  \bibfield  {author} {\bibinfo {author} {\bibfnamefont {A.}~\bibnamefont
  {Schwartz}}, \bibinfo {author} {\bibfnamefont {M.}~\bibnamefont {Dressel}},
  \bibinfo {author} {\bibfnamefont {G.}~\bibnamefont {Gr\"uner}}, \bibinfo
  {author} {\bibfnamefont {V.}~\bibnamefont {Vescoli}}, \bibinfo {author}
  {\bibfnamefont {L.}~\bibnamefont {Degiorgi}}, \ and\ \bibinfo {author}
  {\bibfnamefont {T.}~\bibnamefont {Giamarchi}},\ }\href@noop {} {\bibfield
  {journal} {\bibinfo  {journal} {Phys. Rev. B}\ }\textbf {\bibinfo {volume}
  {58}},\ \bibinfo {pages} {1261} (\bibinfo {year} {1998})}\BibitemShut
  {NoStop}%
\bibitem [{\citenamefont {Tsuchiizu}\ \emph {et~al.}(2001)\citenamefont
  {Tsuchiizu}, \citenamefont {Yoshioka},\ and\ \citenamefont
  {Suzumura}}]{Tsuchiizu01}%
  \BibitemOpen
  \bibfield  {author} {\bibinfo {author} {\bibfnamefont {M.}~\bibnamefont
  {Tsuchiizu}}, \bibinfo {author} {\bibfnamefont {H.}~\bibnamefont {Yoshioka}},
  \ and\ \bibinfo {author} {\bibfnamefont {Y.}~\bibnamefont {Suzumura}},\
  }\href@noop {} {\bibfield  {journal} {\bibinfo  {journal} {J. Phys. Soc.
  Jpn.}\ }\textbf {\bibinfo {volume} {70}},\ \bibinfo {pages} {1460} (\bibinfo
  {year} {2001})}\BibitemShut {NoStop}%
\bibitem [{\citenamefont {Hlubina}\ and\ \citenamefont
  {Rice}(1995)}]{Hlubina95}%
  \BibitemOpen
  \bibfield  {author} {\bibinfo {author} {\bibfnamefont {R.}~\bibnamefont
  {Hlubina}}\ and\ \bibinfo {author} {\bibfnamefont {T.~M.}\ \bibnamefont
  {Rice}},\ }\href@noop {} {\bibfield  {journal} {\bibinfo  {journal} {Phys.
  Rev. B}\ }\textbf {\bibinfo {volume} {51}},\ \bibinfo {pages} {9253}
  (\bibinfo {year} {1995})}\BibitemShut {NoStop}%
\end{thebibliography}%

\end{document}